\RequirePackage{lineno}
\documentclass[a4paper,12pt]{article}
\usepackage{slashed}
\usepackage[affil-sl,auth-sc]{authblk}
\usepackage{amssymb,amsmath,amsbsy}
\usepackage{bbm}
\usepackage{mathrsfs}
\usepackage{mathbbol}
\usepackage{bm}
\usepackage{appendix}
\usepackage{hyperref}
\usepackage[top=1.2in, bottom=1.2in,left=0.9in,includefoot]{geometry}

\usepackage{graphicx}
\usepackage{cite}
\usepackage{float}
\usepackage{caption}
\usepackage{wasysym}
\usepackage{amsthm}
\usepackage[utf8]{inputenc}
\usepackage[english]{babel}
\newif\ifContLineTwo
\newif\ifContLineThree
\def\conC#1{\vbox{\ialign{##\crcr
  \ifContLineThree\hrulefill\else\vphantom{\hrulefill}\fi\crcr
  \noalign{\kern3.2pt\nointerlineskip}
  \ifContLineTwo\hrulefill\else\vphantom{\hrulefill}\fi\crcr
  \noalign{\kern3.2pt\nointerlineskip}
  \ifContLineOne\hrulefill\else\vphantom{\hrulefill}\fi\crcr
  \noalign{\nointerlineskip}
  $\hfil\textstyle{\vbox to 14pt{}#1}\hfil$\crcr}}}
\def\DrawLeg#1#2{
  \kern-.2pt              
  \dimen2 =#1             
  \advance\dimen2 by 2pt  
  \dimen3 = 10.6pt        
  \dimen4 =3.6pt          
  \advance\dimen3 by -\dimen2 
  \multiply\dimen4 by #2
  \advance\dimen3 by \dimen4
  \raise\dimen2 \hbox{\vrule height\dimen3 width .4pt} 
  \kern-.2pt}             
\def\begC#1#2{\setbox0 =\hbox{$\textstyle{#2}$}
  \dimen0=.5\wd0 \dimen1=\ht0
  \conC{\hskip\dimen0}
  \count255=#1
  \ifnum\count255 =1 \ContLineOnetrue\else
  \ifnum\count255 =2 \ContLineTwotrue\else
  \ifnum\count255 =3 \ContLineThreetrue\fi\fi\fi
  \DrawLeg{\dimen1}{\count255}
  \conC{\hskip\dimen0}
  \kern-\dimen0\kern-\dimen0 \box0}
\def\endC#1#2{\setbox0 =\hbox{$\textstyle{#2}$}
  \dimen0=.5\wd0 \dimen1=\ht0
  \conC{\hskip\dimen0}
  \count255=#1
  \ifnum\count255 =1 \ContLineOnefalse\else
  \ifnum\count255 =2 \ContLineTwofalse\else
  \ifnum\count255 =3 \ContLineThreefalse\fi\fi\fi
  \DrawLeg{\dimen1}{\count255}
  \conC{\hskip\dimen0}
  \kern-\dimen0\kern-\dimen0 \box0}

\newcommand {\signalgbth}   {\ensuremath{g b \rightarrow t {H^{-}} }}
\newcommand {\signalggtbh}  {\ensuremath{g g \rightarrow t \bar{b} \text{H}^-}}

\newcommand {\ttbar}        {\ensuremath{ t \bar{t}}}
\newcommand {\ttbarbbbar}   {\ensuremath{ t \bar{t} {b} \bar{b}}}

\newcommand {\Dcthb}        {\ensuremath{ {t} \rightarrow {{H}^{+}} {b} }}
\newcommand {\Brthb}        {\ensuremath{ \text{Br} \left( {\Dcthb} \right) }}

\newcommand {\Dchtanu}      {\ensuremath{ {{H}^{+}} \rightarrow {\tau}^{+} \nu_{\tau} }}
\newcommand {\Brhtanu}   {\ensuremath{ \text{Br} \left( {\Dchtanu} \right) }}

\newcommand {\Dchtb}        {\ensuremath{ {{H}^{+}} \rightarrow {t} \bar{b} }}
\newcommand {\Brhtb}        {\ensuremath{ \text{Br} \left( {\Dchtb} \right) }}

\newcommand{\bea}{\begin{eqnarray}}
\newcommand{\eea}{\end{eqnarray}}

\usepackage{color}

\newcommand {\pythia} {{\tt PYTHIA8-8.2.26 (PYTHIA8)}}
\newcommand {\pyth}{\tt PYTHIA8}
\newcommand {\madgraph} {{\tt MadGraph\_aMC@NLO-2.6.1 (MG5)}}
\newcommand {\mg} {\tt MG5}

\definecolor{orange}{rgb}{0.9,0.2,0}
\definecolor{brown}{rgb}{0.7,0.3,0.2}
\definecolor{fuxia}{rgb}{1,0,1}
\definecolor{skyblue}{rgb}{0,0.1,0.9}
\definecolor{violetred}{rgb}{0.8,0.13,0.56}
\definecolor{deeppink}{rgb}{1.00,0.08,0.5}
\definecolor{pink}{rgb}{1.00,0.75,0.80}
\definecolor{orchid}{rgb}{0.85,0.44,0.84}
\definecolor{lightpink}{rgb}{1.00,0.71,0.76}
\definecolor{bluish}{rgb}{0,0.6,0.8}
\numberwithin{equation}{section}
\title{\bf Probing Heavy Charged Higgs Boson at the LHC}
\author{\small Monoranjan Guchait \thanks{guchait@tifr.res.in}~ and Aravind H. Vijay \thanks{aravind.vijay@tifr.res.in}}
\affil{\small Department of High Energy Physics, \\
Tata Institute of Fundamental Research, \\
 Homi Bhabha Road, Mumbai-400005, India}

\newcommand{\chh} {\ensuremath{{H}^{\pm}}}

\def\invfb{\text{fb}^{-1}}
\newcommand{\mch}{\ensuremath{m_{{H}^{\pm}}}}

\def\t1 {\widetilde {t_1}}
\def\N1{\widetilde \chi_1^0}
\def\N2{\widetilde \chi_2^0}
\def\N3{\widetilde \chi_3^0}
\def\N0{\widetilde \chi^0}
\def\C1{\widetilde \chi_1^{\pm}}
\def\mst1 {m_{\t1}}
\def\br {\begin{eqnarray}}
\def\er {\end{eqnarray}}

\date{}

\begin{document}
\maketitle
\begin{abstract}
Signature of heavier charged Higgs boson, much above the top
quark mass, is investigated at the LHC Run 2 experiments, 
following its decay mode via top and bottom quark focusing on both 
hadronic and semi-leptonic signal final states. 
The generic two Higgs doublet model framework is considered with a special 
emphasis on supersymmetry motivated Type II model. The signal is found to be
heavily affected by huge irreducible backgrounds due to the 
top quark pair production and QCD events. The jet substructure technique is 
used to tag moderately boosted top jets in order to reconstruct charged 
Higgs mass. The simple cut based analysis is performed optimizing various 
kinematic selections, and the signal sensitivity is found to be 
reasonable for only lower range of charged Higgs masses corresponding to 
3000~fb$^{-1}$ integrated luminosity. However, employing the 
multi-variate analysis(MVA) technique, a remarkable improvement in 
signal sensitivity is achieved. We find that the 
charged Higgs signal for the mass range about $300-600$ GeV is observable 
with 1000~$\invfb$ luminosity. However, for high luminosity,
${\cal L} = 3000 \text{fb}^{-1}$, the discovery potential can be extended to $700-800$ GeV. 
\end{abstract}
\vskip .5 true cm
\newpage
\section{Introduction}
The recent discovery of the 125 GeV Higgs 
boson~\cite{Aad:2012tfa,Chatrchyan:2012xdj} at the CERN Large Hadron 
collider (LHC) provides 
the last missing piece of the Standard Model (SM), and open up a new window 
to explore the physics beyond standard model(BSM). Although, the current 
precision measurements of various properties of the Higgs boson, in 
particular the couplings with fermions and gauge bosons, indicate that it 
is indeed the candidate for the SM Higgs~\cite{Khachatryan:2016vau}, 
nonetheless, it does not rule out many BSM scenario.
Among the plethora of BSM candidates. The supersymmetry based models, 
such as minimal supersymmetric standard model(MSSM) is the most popular 
and very well studied BSM scenario, it provides elegant solutions to some of 
the short comings of the SM, and predicts rich and diverse phenomenology
to be testable directly in colliders.

Recall, the MSSM requires at least two Higgs doublets to make 
the theory anomaly free, and also to generate the masses of up and down 
type of fermions. The theories with extended Higgs sector predict
more Higgs boson - neutral and charged states. 
In general, two Higgs doublet model(2HDM) consisting of an extra SU(2) 
Higgs doublet added with the SM Higgs doublet, is well motivated 
and consistent with the Higgs discovery. In fact, the 2HDM can be 
interpreted as the effective theory at low energy of many BSM 
theories with UV completion. For example, the Higgs sector in supersymmetric 
model may appear
as a simple 2HDM (Type II), if the masses of all
sparticles decouple at a very high scale. Generally, 
2HDM is classified into four categories, Type I, II, III and IV  
depending on the nature of Yukawa couplings,
subject to $Z_2$ symmetry in order to avoid Flavor 
changing neutral current
(For more details about 2HDM, see Ref.~\cite{Gunion:425736} and ~\cite{Branco:2011iw}). 
In all classes of 2HDM scenario, there exist five physical 
Higgs boson states, two CP even ($h, H$, with the assumption, $m_h < m_H$),
one CP odd ($A$), and two 
charged Higgs bosons($\chh$). The lightest CP even Higgs $h$ can be interpreted
as the SM-like Higgs boson in the decoupling limit, where the other 
states turn out to be very heavy, much above the electroweak 
scale \cite{Gunion:2002zf}. However, some other studies also show 
that CP even Higgs states may behave as SM-like with mass 125 GeV 
in the alignment limit even without 
decoupling \cite{Carena:2013ooa,Dev:2014yca,Carena:2014nza,Profumo:2016zxo}.
Presence of extra physical Higgs boson states along 
with the SM-like Higgs is one of the characteristics of BSM. 
Needless to say, discovery of an extra
Higgs boson certainly confirms the existence of BSM.
Therefore, looking for these additional Higgs bosons 
in various decay channels over a wide range of masses is
a top priority program in the current LHC experiment.

In this context, searching for the charged Higgs boson signal is unique, 
since discovery of it clearly, and unambiguously confirms the presence 
of BSM. Therefore, the study of charged Higgs boson has received 
special attention both phenomenologically and experimentally.
For the lower mass range, less than the top quark mass, $\mch < m_t$,
the phenomenology of the charged Higgs boson is well studied, 
and also experimentally probed thoroughly in many of its 
decay channels. However, detection of the charged Higgs boson for the 
heavier mass range, greater than the top quark mass($m_{\chh} \gg m_t $), 
is found to be very challenging due to huge contamination by the irreducible  
SM backgrounds.
In this current study, we attempt to find the discovery potential of
the charged Higgs boson for this heavier mass range ($\mch \gg m_t$).
The study is carried out within the framework of the generic 2HDM
with an emphasis on Type II 2HDM motivated by supersymmetry.
The charged Higgs boson couplings with fermions are strongly dependent
on $\tan\beta$, and hence its production and subsequent decays  
sensitive to $\tan\beta$. In hadron colliders, in the lower mass 
range $\left( {m_{\chh}} < m_t \right)$, the charged Higgs bosons are 
produced via a pair production of top quark, $p\bar p$/$pp \to t\bar t$ 
following the decay {\Dcthb}. For intermediate and heavier mass range,
it is mainly produced directly in association with a top 
quark(and also a $b$ quark)~\cite{deFlorian:2016spz}. Furthermore,
charged Higgs boson can be produced in SUSY cascade decays via heavier
chargino and neutralino production in gluino and
squark decays~\cite{Datta:2001qs,Datta:2003iz}.
 
So far, non observation of any charged Higgs signal events in direct searches
constrain its production and decay in a model independent way,
which in turn can be translated to exclude the relevant parameter space,
in particular $\tan\beta$ and $m_{\chh}$, for a given model framework.
For example, in the past, direct searches at LEP \cite{Abbiendi:2013hk} and
Tevatron \cite{Aaltonen:2009ke} experiments
excluded lower mass range of $m_{\chh}$ in terms of $\tan\beta$.
At the LHC Run 1 experiments with $\sqrt{s}=7$ and 8 TeV data, 
lighter charged Higgs boson is probed in the decay channels
$\tau\nu$ \cite{Aad:2014kga,Khachatryan:2015qxa}, 
$cs$ \cite{Aad:2013hla,Khachatryan:2015uua} and also 
$cb$\cite{CMS:2016qoa}, while at Run 2 with $\sqrt{s}=13$ TeV energy, mainly 
the decay modes $\tau\nu$ \cite{CMS:2016szv,Aaboud:2016dig}, and 
$tb$ \cite{ATLAS:2016qiq} are considered to probe it upto $\sim 1$ TeV mass.  
The absence of any signal event in ${\Dchtanu}$ decay modes
in CMS at 13 TeV energy with an integrated luminosity 12.9 $\invfb$ 
leads an 
exclusion of the cross section times respective branching ratio 
for the mass range $180 \text{ GeV} < m_{\chh} < 3 \text{ TeV}$, 
where as limits on the ${\Brthb} \times {\Brhtanu}$ 
are set for the range $80 \text{ GeV} <  m_{\chh} < 160 \text{ GeV}$ \cite{CMS:2016szv}.
Eventually, these exclusion limits, rule out 
$m_{\chh}\sim 90-160$~GeV corresponding to the entire range 
of $\tan\beta$ up to 60 in the context of MSSM with $m_h^{mod+}$ 
scenario~\cite{CMS:2016abv}, 
except a hole around $m_{\chh}\sim150-160$, and $\tan\beta\sim10$.
Similar results are also published from 
ATLAS \cite{Aaboud:2016dig} at $\sqrt{s}=13$~TeV.
The searches in the {\Dchtb} decay channel for heavier mass range carried 
out by ATLAS at $\sqrt{s}=13$ TeV and ${\cal L}=13.2$ fb$^{-1}$ 
excluded $\mch \sim 300 - 900$ GeV for a very 
low $\tan \beta \left(\sim 0.5 - 1.7 \right )$ region \cite{ATLAS:2016qiq}, 
where as for high values of $\tan\beta >44(60)$, $\mch \sim 300$(366)GeV 
are excluded. Note that this decay channel is also probed at 
$\sqrt{s}=8 $ TeV by ATLAS including the s-channel charged Higgs 
production, and exclusion are presented for the cross section times 
Br$\left(\Dchtb\right)$\cite{Aad:2015typ}.
However these limits are found to be  
very weak in comparison to the predictions from $\Dchtanu$ 
searches \cite{Aad:2014kga,Aaboud:2016dig}.
Remarkably, the most stringent constraints on the charged Higgs sector in 
the context of SUSY motivated Type II type of model are predicted 
indirectly by the neutral 
Higgs boson searches, $ p p \to {h}, {H}, {A} \to \tau \bar{\tau}$
at the LHC~\cite{Sirunyan:2018zut}. It can be attributed to the fact that the 
neutral Higgs boson couplings with tau leptons very strongly depends on 
$\tan\beta$, in particular for higher values of it. 
Exclusion region predicted by these neutral Higgs boson 
searches, imply a limit on $\tan\beta>6$ for $m_A <250$~GeV, where as
Higher values of $\tan\beta$($> 20 $ are completely ruled out up to 60,
for $m_A \sim m_{\chh} \sim 1000$~GeV, $m_A$ is the mass 
of the pseudoscalar Higgs, related 
with the charged Higgs mass as, $\mch^2 = m_W^2 + m_A^2$
in SUSY model (like Type II model).
In addition to these direct limits, 
the charged Higgs sector is also constrained by 
flavor physics data. Strong contribution via loops to the Br of
rare decay modes of B meson makes it very sensitive to flavor physics
observables. Measurements of these Br by B-factories, and also 
at the LHC and LHCb put very strong limit to the charged Higgs sector.
More details about these latest constraints in the framework of 2HDM can 
be found in a recent review of Ref.~\cite{Arbey:2017gmh}, 
and references therein.

In the phenomenological side, there have been numerous
studies on exploring the $\chh$ signal in various decay
channels in the context of the MSSM Higgs sector
\cite{Krawczyk:2017sug,Arhrib:2017wmo,Basso:2013wna,Arhrib:2016wpw,Assamagan:2004gv,Roy:1999xw,Moretti:1996ra,Moretti:2016jkp,Enberg:2014pua,Guedes:2012eu,DiazCruz:2007tf}, and as well as in 2HDM framework 
\cite{Moretti:2016qcc,Basso:2013wna,Aoki:2011wd,Basso:2012st} 
 using various interesting techniques. 
More details about charged Higgs phenomenology can be found in 
Ref.~\cite{Akeroyd:2016ymd}. 
It is worth to mention here 
about the use of
$\tau$ lepton polarization in its 1 and 3 prong decay
for {\Dchtanu}, which is found to be very useful
in extracting the signal suppressing $t \bar{t}$ and QCD background
\cite{Raychaudhuri:1995cc,Guchait:2008ar,Guchait:2006jp}.
The signal of charged Higgs boson is also probed in the subdominant production
channels ${\chh} W^\mp$ \cite{Moretti:1998xq} and 
${H}^+ {H}^-$ \cite{BarrientosBendezu:1999gp,Krause:1997rc}.
Discovery potential of charged Higgs for heavier
mass (${\mch} > m_t$) range with its dominant decay mode 
{\Dchtb} is investigated by many authors in the framework of
SUSY models~\cite{Gunion:1993sv,Barger:1993th,Miller:1999bm,Moretti:1999bw}.
For instance, in the Ref.\cite{Moretti:1999bw}, authors used
triple and four b-tagging in order to suppress SM background,
which also costs signal significantly as well.
Consequently, for heavier mass range, it is found 
to be very hard to achieve a reasonable signal sensitivity, 
due to large $t\bar t$ and QCD backgrounds.
A recent study~\cite{Yang:2011jk} reports about the detection prospect 
of charged Higgs signal for heavier mass $\gtrsim 1$ TeV 
applying jet substructure technique to tag top quarks from the charged 
Higgs decay in the framework of 2HDM. The authors predicted 
reasonable sensitivities of charged Higgs signal around the mass of 
1~TeV, and found difficult to probe it for the intermediate mass range.
The jet substructure technique is also used to look for
heavy charged Higgs boson signal in the decay 
channel $\chh \to W^\pm A$ for lighter
$A$ boson states\cite{Patrick:2016rtw,Pedersen:2016kyw}.
In this current study, we explore the detection prospect 
of charged Higgs boson for the intermediate to heavier 
mass range, $300 - 1000$ GeV, considering the decay mode,
{\Dchtb} with the hadronic and leptonic final state.
For heavier mass of $\chh$, the top
quark from its decay is expected to be boosted, and 
we try to exploit this feature by employing the technique 
of jet substructures to reconstruct the top quark, 
and subsequently the charged Higgs boson. 
This method helps to avoid 
the combinatorial problem while reconstructing the top quark 
simply by combining the hard jets. In this
study, first we attempt to obtain signal sensitivity using  
cut based analysis, and then try to improve the sensitivity employing 
the multivariate(MVA) analysis. 
Performing a detail analysis in the MVA
framework, we achieve a remarkable improvements in signal
sensitivity, and
results are presented for three integrated luminosity options
${\cal L}=300$ $\invfb$, 1000 $\invfb$ and 3000 $\invfb$. Finally, 
for the sake of completeness,
signal sensitivities are predicted for all classes of 2HDM corresponding 
to a few 
benchmark parameter space.    

We present this study as follows. Briefly describing the 2HDM in 
Sec.\ref{modelref}, the charged Higgs production is discussed in 
Sec.\ref{productionofchh}. In Sec.\ref{sec4}, the signal 
and backgrounds are discussed, and subsequently, details of simulation
are presented in the subsection \ref{simulation} with a brief 
description about top tagging in subsection \ref{toptagging}. 
The results based on cut and count analysis are discussed in
subsection \ref{cutbasedresults}, while in Sec.\ref{MVAResults}
the results based on MVA analysis are presented. Finally we summarize in
Sec.\ref{summary}.        

\section{Two Higgs doublet Model}
\label{modelref}
\begin{figure}[t]
	\begin{center}
		\caption{Charged Higgs branching ratios for four classes of 
			2HDM, setting $m_H=m_A=\mch=500$ GeV and $\sin(\beta-\alpha)=$1.}
		\label{fig:BRCH}
		\begin{tabular}{cc}
			\includegraphics[width=0.46\textwidth]{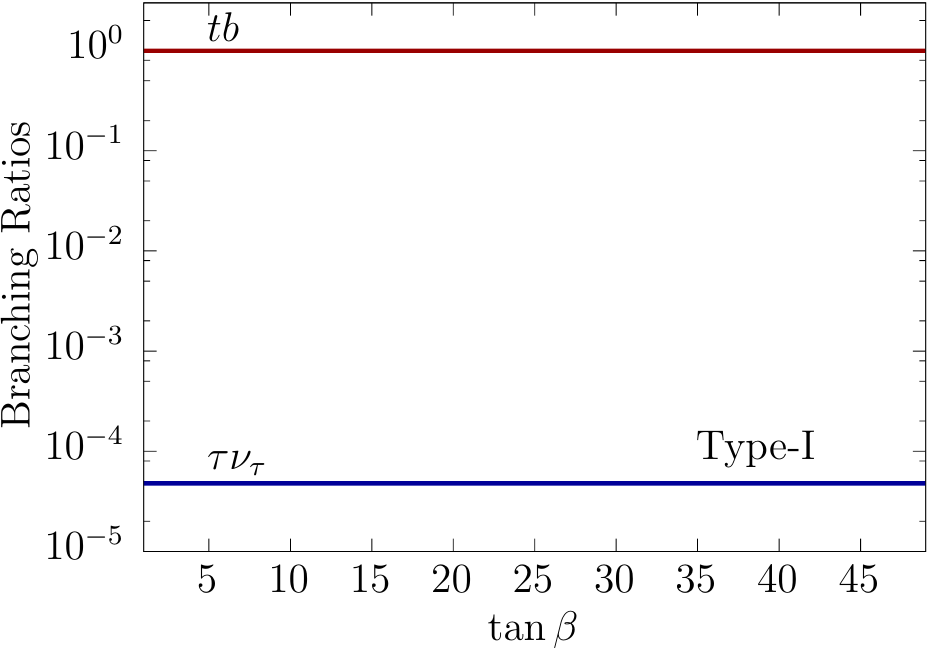} &
			\includegraphics[width=0.46\textwidth]{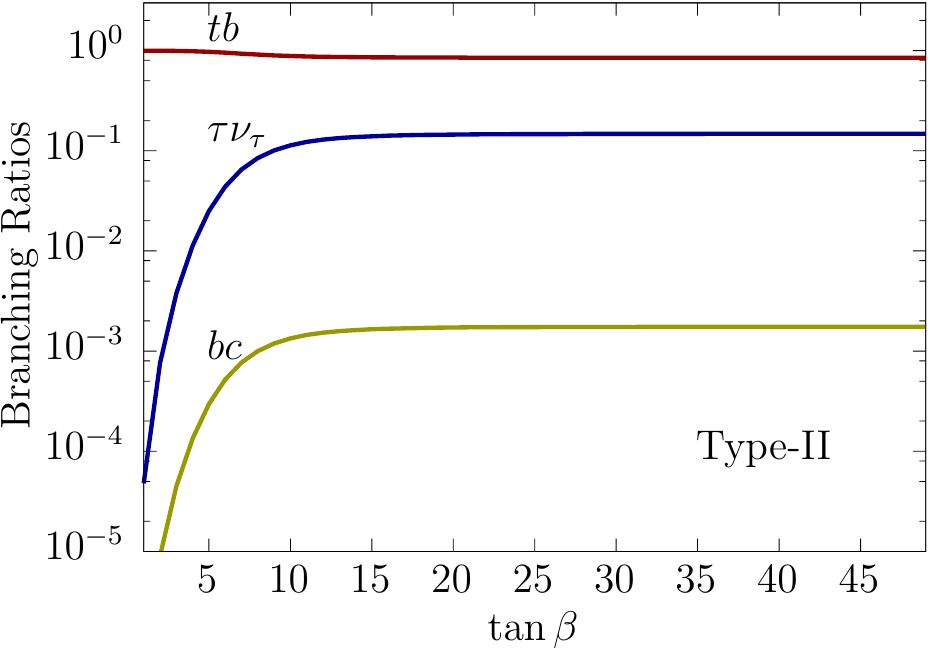}  \\
			\includegraphics[width=0.46\textwidth]{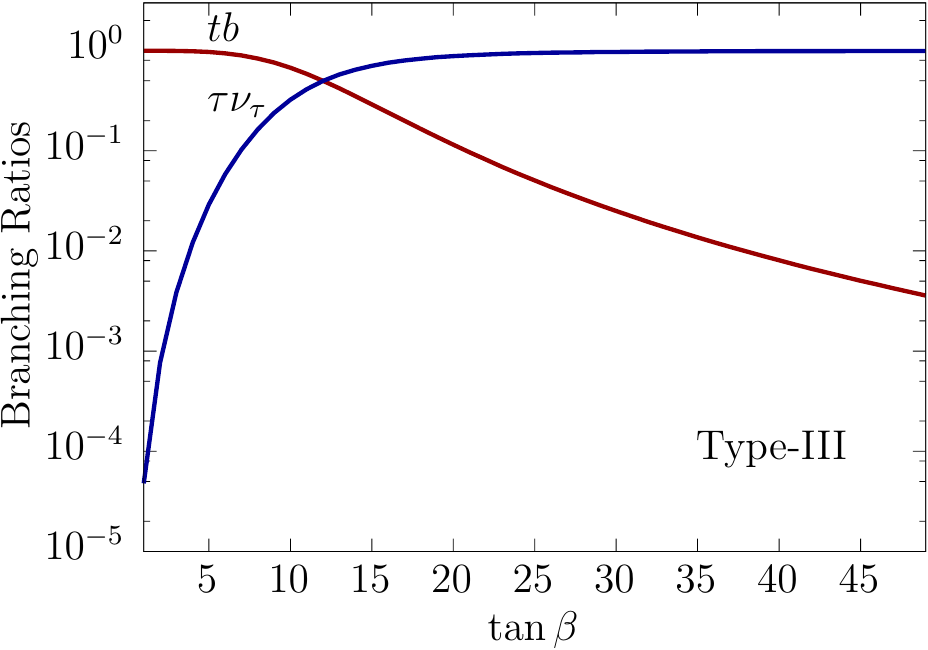} &
			\includegraphics[width=0.46\textwidth]{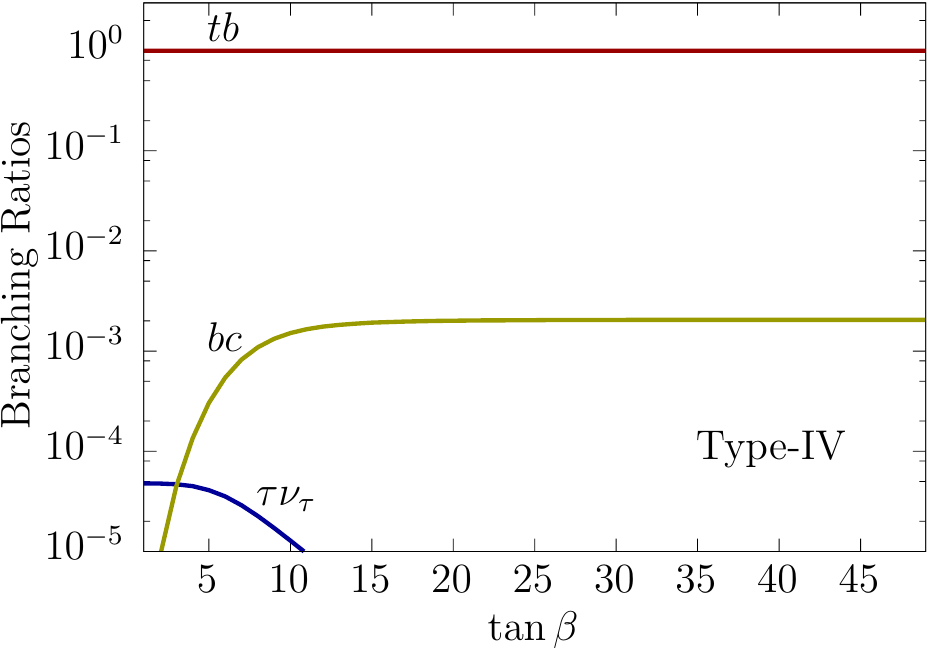}  \\
		\end{tabular} 
	\end{center}
\end{figure}

In the context of our present study, it is instructive to discuss very 
briefly about the 2HDM.
In this model, an extra SU(2) Higgs
doublet is added with the SM Higgs 
doublet. 
The most general 2HDM potential consisting 
two doublets $\phi_1$ and $\phi_2$ with 
hypercharge Y=+1 is given by~\cite{Gunion:425736,Branco:2011iw},
\br
V &=& m_{11}^2 \phi_1^2 + m_{22}^2 \phi_2^2 
    - m_{12}^2(\phi_1^\dag \phi_2 +\phi_1 \phi_2^\dag) \nonumber \\
  &+& \frac{\lambda_1}{2}(\phi_1^\dag\phi_1)^2 + \frac{\lambda_2}{2}
          (\phi_2^\dag\phi_2)^2 + \lambda_3(\phi_1^\dag\phi_1)
           (\phi_2^\dag\phi_2) \nonumber \\
  &+& \lambda_4(\phi_1^\dag\phi_2)(\phi_2^\dag\phi_1) + 
      \frac{\lambda_5}{2} \left [ \left (\phi_1^\dag \phi_2)^2 + 
           (\phi_2^\dag\phi_1)^2  \right ) \right ]  
\er     
For simplifications, all the free parameters are assumed to be real to 
conserve CP property, and the discrete $Z_2$ symmetry, 
$\phi_1 \to -\phi_1$  and 
$\phi_2 \to +\phi_2$ is imposed to suppress FCNC at the tree level. 
The $Z_2$ symmetry is softly broken by the terms proportional to
$m_{12}$.
The minimum of the potential $V$ is ensured by two vacuum expectation 
values(vevs), which break the symmetry down to U(1)$_{\text{em}}$ 
symmetry,
\br
<\phi_1> = \left [ {\begin{array}{c}
0 \\ 
\frac{v_1}{\sqrt{2}} \\   
\end{array} } \right ],
<\phi_2> = \left[ {\begin{array}{c} 
0  \\
\frac{v_2}{\sqrt{2}} \\ 
\end{array} } \right ]
\er
where $v_1$ and $v_2$ are two vevs corresponding to neutral components
of $\phi_1$ and $\phi_2$ respectively, 
with $v = \sqrt{v_1 ^2 + v_2^2}$. The ratio of two vevs defined to be
$\tan\beta = \frac{v_2}{v_1}$ is considered as one of the free parameter 
of the model. Expanding the doublets around the minimum of the potential, 
the Higgs fields 
can be given by\cite{Gunion:425736,Branco:2011iw},
\br
\phi_1 = \left ( {\begin{array}{c}
\phi_1^+ \\
\frac{1}{\sqrt{2}}(v\cos\beta + \phi_1^0) \\ 
\end{array} } \right ),
\phi_2 = \left ( {\begin{array}{c}
\phi_2^+ \\  
\frac{1}{\sqrt{2}}(v\sin\beta + \phi_2^0). \\
\end{array} } \right )
\er
Already mentioned in the previous section that 
after symmetry breaking, the potential predicts five physical 
Higgs boson states, two neutral CP even states, $h$ and $H$($m_h< m_H$), 
one neutral CP odd state A, and two charged 
states $\chh$. The physical charged
state and CP odd neutral states are  expressed as,
\br
{\chh} = -\phi_1^\pm \sin\beta + \phi_2^\pm \cos\beta, \\
A = \sqrt{2}\left ( - \text{Im} \phi_1^0 \sin\beta + \text{Im} \phi_2^0\cos\beta \right ). 
\er
The two CP even neutral weak states mix through an angle $\alpha$ providing  
two mass eigenstates, $h$ and $H$. The input parameters present in the potential 
V, can be re-expressed in terms of physical masses and other parameters such
as,
\br
m_h, m_H, m_A, {\mch}, \tan\beta, \sin(\beta-\alpha), v,  m_{12}^2 . 
\er
Note that $v$ is set to be at the electroweak scale ($=246$ GeV), and one 
of the CP even Higgs boson can be interpreted as the recently discovered 
Higgs boson of mass 125 GeV under certain scenario of the model, which are
already mentioned in the earlier section
\cite{Gunion:2002zf,Carena:2013ooa,Dev:2014yca,Carena:2014nza,Profumo:2016zxo}.
The topics of our interest in this current study is to look for the
charged Higgs signal, hence we focus only this sector of 2HDM.  
In the generic 2HDM model,
the Yukawa couplings of charged Higgs with fermions are given
by \cite{Gunion:425736,Branco:2011iw},

\begin{eqnarray}
    {\cal L}_{H^{\pm}} & = & -H^{+}\left[\frac{\sqrt{2}V_{ud}}{v}\bar{u}\left(m_{u}\lambda_{u}P_{L}+m_{d}\lambda_{d}P_{R}\right)d+\frac{\sqrt{2}m_{\ell}}{v}\lambda_{\ell}\bar{v}_{\ell}\ell_{R}\right]+\text{H.C}
\label{eq:Hcoupl}
\end{eqnarray}
where $V_{ud}$ is the CKM matrix elements, and the couplings 
$\lambda$s represent either $\tan\beta$ or $\cot\beta$ depending on the 
assignments of $Z_2$ charges to right handed fermions, which finally define
the four types of 2HDM. The Table~\ref{tab:TypeCoupling} 
presents $\lambda$s corresponding to four types of 2HDM model. 
\begin{table}
    \begin{center}
    	\caption{$\lambda$s in charged Higgs couplings with fermions(Eq.~\ref{eq:Hcoupl}) for all four types of 2HDM.}
        \label{tab:TypeCoupling}
        \begin{tabular}{ccccc}
            \hline
            & Type I & Type II & Type III & Type IV \\
            \hline
            $\lambda_u$    & $\cot\beta$ & $\cot\beta$  & $\cot\beta$  & $\cot\beta$  \\
            $\lambda_d$    & $\cot\beta$ & $-\tan\beta$ & $\cot\beta$  & $-\tan\beta$ \\
            $\lambda_\ell$ & $\cot\beta$ & $-\tan\beta$ & $-\tan\beta$ & $\cot\beta$  \\
            \hline 
        \end{tabular}
    \end{center}
\end{table}
As shown 
in Type I model, the couplings of charged Higgs 
with fermions are heavily suppressed for $\tan\beta \gg 1$, 
same as
Type III model, 
except the coupling with lepton which is enhanced making it lepton specific. 
In the Type II model which is same as the 
supersymmetric Higgs sector, couplings are favored with u-type
quarks for low $\tan\beta$ case, 
where as for d-type quarks and leptons, high values 
of $\tan\beta$ are preferred.
The Type IV model is found to be lepto-phobic for high $\tan\beta$ scenario
but, the couplings with quarks are same for both Type II and Type IV 
model.
Consequently, the charged Higgs decay Br to 
fermions are very much $\tan\beta$ dependent. 
The decay channels of  charged Higgs to 
$\tau\bar{\nu}_{\tau}$ or $\bar{ t} { b}$ channels are very much 
sensitive to $\tan\beta$ once they are kinematically allowed.
The charged Higgs Br computed by HDECAY~\cite{Djouadi:2018xqq,Djouadi:1997yw}
are demonstrated for various values of $\tan\beta$
setting $\mch=500$ GeV, in Fig.\ref{fig:BRCH} 
corresponding to four 
Types of 2HDM. The input parameters are set as,
$m_h=125$ GeV, $m_H=m_A=m_{\chh}$ and $\sin(\beta-\alpha)=1$, like 
MSSM scenario~\cite{Arbey:2017gmh} with decoupling limit. 
In Type I model, due to the $\cot\beta$ dependence of coupling, 
the {\Brhtanu} is suppressed by $m_\tau^2/m_t^2$ 
over {\Brhtb}, leading almost 100\% Br to $\bar t b$ mode.
The dominant decay mode of charged Higgs in Type II model, as expected  
is in the $\bar{t} {b}$ channel, following sub-dominant $\bar{\tau}\nu_{\tau}$ 
channel with Br $~\sim 10-15$ \%,
followed by other suppressed
modes such as, ${H}^+ \to \bar{b} {c}, {c} \bar{s}$. However, in the case of 
Type III model, 
which is lepton specific, the charged Higgs decays to     
$\bar{\tau}\nu_{\tau}$ mode dominantly, except in the lower region of 
$\tan\beta \sim 1-12$ where $ t \bar b$ mode becomes important. 
On contrary, $\bar{\tau}\nu_{\tau}$ mode gets suppressed in Type IV model, 
because of $\cot\beta$ dependence, and
$\bar t b$ channel takes over. It is to be noted that the pattern of these 
Brs expected to be different in the presence 
of $\chh \to {W}^\pm \phi,\ \ \left( \phi={h},{H},{A} \right)$ mode, 
of which decay width 
is proportional to $\cos \left( \beta-\alpha \right) $ leading 
it dominant ($\sim 100 \%$) for the choice of 
$\sin(\beta-\alpha)=0$ \footnote{This 
scenario is equivalent to MSSM inverted scenario where $H$ is the SM-like and
$m_H=125 $\cite{Arbey:2017gmh}}. 
Interestingly, in the case of SUSY motivated Higgs sector, i.e
in Type II model, if kinematically allowed, 
the charged Higgs can decay also to chargino and neutralino 
pair, $\chh \to \tilde\chi^\pm_i \tilde\chi_j^0;$ (i:1-2, j:1-4), which 
may be dominant for Higgsino like scenario~\cite{Bisset:2000ud}.
As pointed out earlier, that the charged Higgs sector is severely constrained
by flavor physics data in addition to the direct searches of which
details can be found in reviews~\cite{Akeroyd:2016ymd,Arbey:2017gmh}.

\section{Charged Higgs production}
\label{productionofchh}
In the intermediate to heavier mass range (${\mch} \gtrsim m_t$), 
the charged Higgs is produced directly in proton-proton 
collision via the process, 
\br
{p p} \to {t} {H}^- + X. 
\er
At the parton level, the production mechanism is initiated via 
two subprocesses, 
\begin{eqnarray}
	{g g}, {q} \bar{q} & ~\rightarrow ~~{t}\bar{b}{H}^{-} & \left(\text{4FS}\right) \nonumber\\
	{g b}              &  \rightarrow ~~{t} {H}^{-}        & \left(\text{5FS}\right) \label{eq:Production}
\end{eqnarray}
in 4 flavor(4FS) and 5 flavor 
scheme(5FS) at the leading order(LO) respectively.
In fact, the process in 4FS is part of the NLO QCD correction to the 
5FS scheme mechanism.
The total NLO QCD effects to the inclusive $\chh$ production 
is essentially the NLO correction to the process ${gb} \to  {t H}^-$ plus the total contribution due to the tree-level processes \cite{Dittmaier:2009np}.
In 5FS, the NLO QCD corrections are known for sometime
in the literature~\cite{Plehn:2002vy,Berger:2003sm, Kidonakis:2004ib, Kidonakis:2005hc},
and also very recently approximate NNLO calculations also
published~\cite{Kidonakis:2016eeu}. The total theoretical uncertainty in $\chh$ production in association with top quark(5FS) is found to be of the 
range 15-20\% \cite{Flechl:2014wfa}.
In the 4FS, the final state bottom quark which originates due to 
the hard scattering is assumed to have non zero mass, where 
as in 5FS, the $b$ quark is treated as massless, and being part 
of the parton flux. In 4FS, 
the corresponding NLO correction estimated to be around 20\% for 
the lower range of $\mch$, and
it goes up little for more higher masses~\cite{Flechl:2014wfa}.

At finite order, the cross section in
4FS does not match with 5FS, as expected, due to different 
ways of treating perturbative calculation.
However, it is expected that the results will match within the respective 
uncertainties 
once taking into account of all orders in perturbation. 
In order to obtain the precise estimation of charged Higgs production 
cross section, one needs to combine the 4 and 5 flavor scheme 
predictions appropriately. This combination is performed following 
the prescription, so called Santander-matching~\cite{Harlander:2011aa}.
In the IR limit ($\frac{m_{\chh}}{m_{b}} \rightarrow 1$),
the cross sections obtained from 
4FS and 5FS scheme match nicely. 
The main difference between the 4FS and 5FS occurs because of
the presence of large logarithm, which arises due to the 
splitting of incoming gluon into two nearly collinear b quarks \cite{Maltoni:2012pa}. Thus, 
the calculated cross sections using two schemes should be combined 
in such a manner that such logarithmic effects are taken into 
account appropriately. 
The prescription to match these cross sections computed in two 
schemes is given by~\cite{deFlorian:2016spz,Flechl:2014wfa},
\br
\sigma = \frac{\sigma^{\text{4FS}} + w \sigma^{\text{5FS}}}{1+w}\ \ {\rm with}
\  \ w = \ln\frac{ \mch }{m_{b}}-2 .          
\label{eq:4FS}
\er
Similarly, the theoretical uncertainties are combined as, 
\br
\Delta \sigma = \frac{\Delta\sigma^{\text{4FS}} + w \Delta \sigma^{\text{5FS}}}{1+w}\ \  {\rm with}\ \ w = \ln\frac{ \mch }{m_{b}}-2.          
\label{eq:5FS}
\er
With this matching methodology, the overall theoretical uncertainty 
of the combined NLO cross section is found to be around 10\%, 
where as the individual 4FS and 5FS 
cross sections at NLO are in reasonable agreement within $\sim$ 20\% 
from the central value \cite{Flechl:2014wfa,deFlorian:2016spz}.  
The production cross section and the corresponding 
uncertainty are very sensitive 
to $\tan\beta$, owing to the dependence of Yukawa coupling on it.  
The scale of uncertainty reduces with the decrease of $\tan\beta$ 
through the correction of bottom Yukawa coupling, which is proportional 
to $\tan\beta$. 
We first estimate the charged Higgs boson production in Type II 2HDM 
motivated by the SUSY providing inputs $\tan\beta$ and $\mch$, 
and then predict the corresponding cross sections
for other classes of 2HDM(Type I, III and IV) simply by 
appropriately rescaling the couplings.
It is to be noted that in the MSSM, the NLO QCD corrections may involve 
additional loop contributions from gluinos and squarks, which
also depend on $\tan\beta$. This extra contribution can be absorbed 
through the rescaling of the the NLO QCD prediction of the bottom Yukawa 
coupling~\cite{Borzumati:1999th}. The total cross section
primarily governed by the ${tb}\chh$ coupling is found to be minimum in 
strength for $\tan\beta \approx 7-8$. 
\begin{table}
	\begin{center}
		\caption{
			Charged Higgs boson production cross
			sections(in fb) in 4FS and 5FS schemes, at $\sqrt{s}=13$~TeV
			and $\tan \beta=30$ in Type II model and the last row presents
			cross sections for Type I. $\mu$ and $m_{b}(\mu)$ represent the QCD
			scales and mass of bottom quark at the scale $\mu$.
		}
		\begin{tabular}{|c|c|c|c|c|c|}
			\hline 
			$\mch$ (GeV)$\to$ & 300  & 500  & 600  & 800  & 1000 \tabularnewline
			\hline 
			$\mu$ (GeV) & 236.5  & 336.5  & 386.5  & 486.5  & 586.5 \tabularnewline
			$m_{b}\left(\mu\right)$ (GeV)  & 2.64  & 2.58  & 2.56  & 2.51  & 2.48 \tabularnewline
			\hline 
			$\sigma\left({ pp}\rightarrow { tbH}^{\pm}\right)$ (4FS) LO & 290.2  & 60.4  & 30.6  & 9.0  & 3.1 \tabularnewline
			$\sigma\left({ pp}\rightarrow { tbH}^{\pm}\right)$ (4FS) NLO & 359.4  & 73.3  & 39.9  & 11.4  & 4.1 \tabularnewline
			$\sigma\left({ pp} \rightarrow { tH}^{\pm}\right)$ (5FS) LO & 581.3  & 126.0  & 64.8  & 19.7  & 6.9 \tabularnewline
			$\sigma\left({ pp}\rightarrow { tH}^{\pm}\right)$ (5FS) NLO & 748.6  & 166.2  & 86.1  & 26.5  & 9.3 \tabularnewline
			\hline 
Matched (NLO) & 625.3  & 140.9  & 74.1  & 22.8  & 8.1 \tabularnewline
			
Matched (NLO) (Type 1) & 2.9  & $0.66$  & $0.35$  & $0.11$  & $3.9\times10^{-2}$ \tabularnewline
			\hline 
		\end{tabular}
		\label{tab:CHcs}
	\end{center}
\end{table}
In Table~\ref{tab:CHcs}, the charged Higgs boson
production cross sections for both schemes, and the final
matched values are presented for few representative 
choices for $\mch$ and $\tan\beta=30$ in Type II model.
The cross sections are computed both at LO and NLO, 
using {\tt MadGraph5-2.6.1}~\cite{Alwall:2014hca},
with the {\tt FeynRules} \cite{Alloul:2013bka} model file uploaded
by authors of \cite{Degrande:2015vpa}. 
We notice that for $\tan\beta=3$, the cross sections go 
down by a factor of $\sim \frac{1}{2}$ in Type II model. 
In calculating these cross sections,
factorization and renormalization scales are set as, 
$\mu^2  = \left( \frac{ {\mch} + m_{t}}{2} \right)^2$, 
as shown in the first row along with the 
value of running b-quark mass\cite{Bednyakov:2016onn}. 
Variation of cross sections are found to be within a range
from ${\cal O}(100)$fb to ${\cal O}(1)$fb corresponding to the 
range {\mch} $\sim$ 300 - 1000 GeV.

In Type I model (see Table~\ref{tab:TypeCoupling}), the charged Higgs boson 
couplings with top and bottom quark go by $\sim (m_{ b} + m_{ t})\cot\beta$.  
The cross sections in Type I model simply 
can be obtained from the values corresponding to Type II model by 
rescaling the Yukawa 
couplings~\cite{Degrande:2015vpa,Flechl:2014wfa,deFlorian:2016spz}. 
The total cross section can be 
parameterized by
$\sigma_{\chh}^{\text{Type II}}\propto {g}_{t}^{2}\sigma_{t}\cot^{2}\beta+g_{b}\sigma_{b}\tan^{2}\beta+g_{t}g_{b}\sigma_{tb}$,
where, $g_{t}$  and $g_{b}$ are the part of the Yukawa couplings 
proportional to top and bottom quark masses respectively. 
Evaluating the contributions 
by setting $m_{t}=0( {\text{i.e }} g_{t}=0)$ and $m_{b}=0( {\text{i.e }} g_{b}=0)$, 
$\sigma_{b}, \sigma_{t}$ and $\sigma_{bt}$ can be
obtained. Thus, the cross sections in Type I model can be estimated 
rescaling each contribution by $\cot\beta$. This prescription works 
in to all orders in QCD, but not appropriate to all orders in the
electroweak corrections~\cite{deFlorian:2016spz}.      
\begin{figure}[t]
	\begin{center}
		\includegraphics[width=0.7\textwidth]{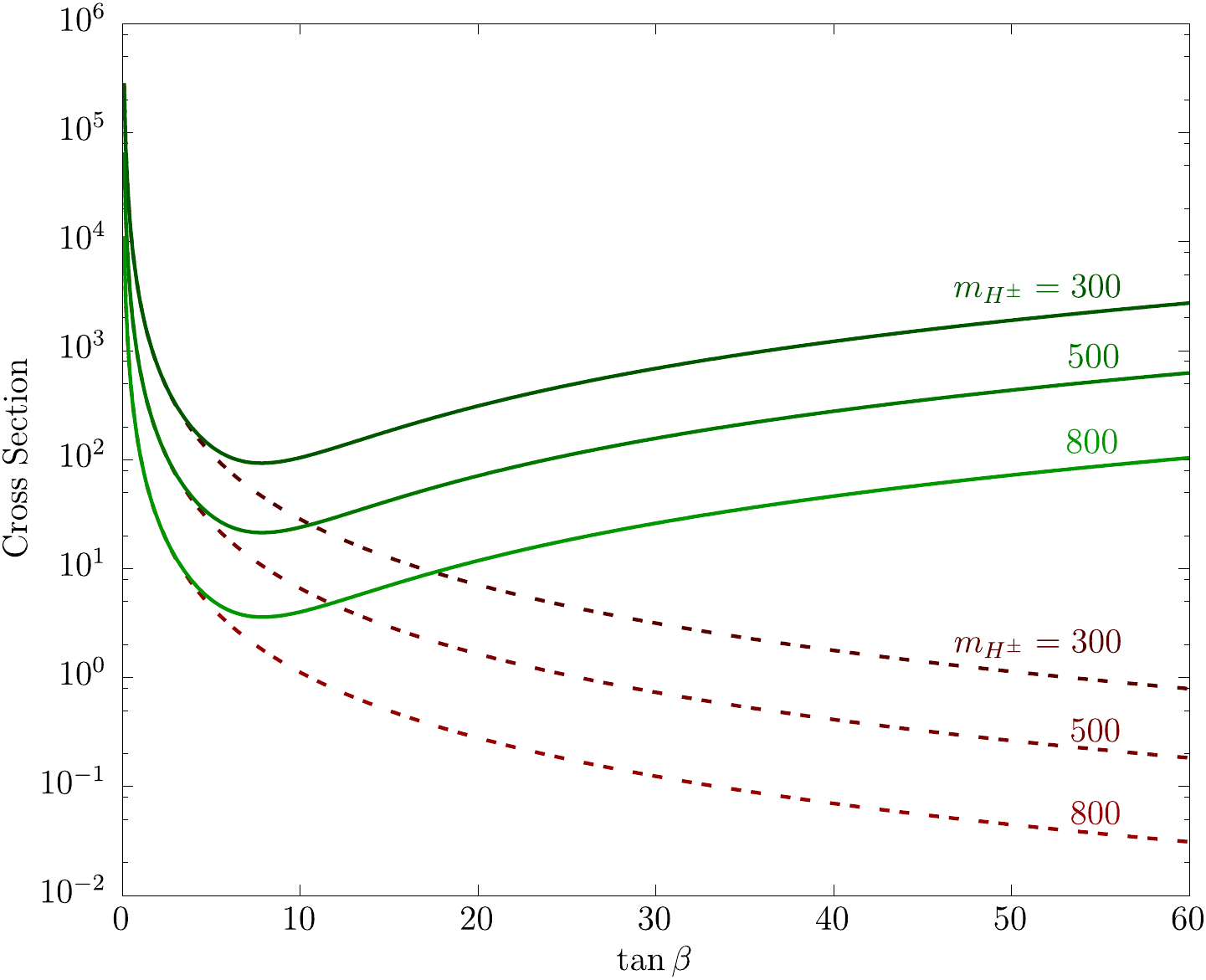}\\
		\caption{Matched charged Higgs production cross section(fb) 
			at $\sqrt{s}=13$ TeV for three different masses(in GeV) in 
			Type II (green/solid) and Type I (red/dashed) 2HDM.}
		\label{fig:CHsigma}
	\end{center}
\end{figure}
The cross sections for both in Type I and II 2HDM are presented in 
Fig.~\ref{fig:CHsigma} for various values of $\tan\beta$ and three choices
of $\mch=300$~GeV, 500~GeV and 800 GeV. Clearly, 
as expected, the cross sections in Type I model are suppressed over Type II 
model by approximately $\sim \tan^2\beta$, for $\tan\beta \gg 1$.
The cross sections in Type III(Type IV) model
are the same as the Type I(Type II) due to the identical Yukawa 
coupling structures with quarks.
A dip is observed for Type II model around 
$\tan\beta \sim 7-8$ unlike Type I, which can be understood 
from the respective couplings dependence on $\tan\beta$ or $\cot\beta$. 
 
\section{Signal and Background}
\label{sec4}
As mentioned before, in this current study, the signature of charged 
Higgs boson is explored with its    
decay mode, {\Dchtb}. 
The {\Brhtb} is almost dominant, more than 70\% for a wide
range of $\tan\beta$, and for all classes of 2HDM  as shown in 
Fig.~\ref{fig:BRCH}, except for the Type III model which is lepton specific.  
Signal is simulated considering $\chh$ production mechanisms, Eq.\ref{eq:Production}, 
and eventually the final results are obtained by combining them following 
the recipe, given in Eq.\ref{eq:4FS}.

The resulting signal final state consists multiple top quarks   
via the following processes:
\begin{equation}
\begin{array}{ccccccc}
5\text{FS} & : & { gb} & \rightarrow & { tH}^{-} & \rightarrow & { t}\bar{ t}{ b}\\
4\text{FS} & : & { gg} & \rightarrow & { t}\bar{b}H^{-} & \rightarrow & { t}\bar{ b}\bar{ t}{ b}
\end{array}\label{eq:sig}
\end{equation}
Both leptonic and hadronic final states are 
considered following the semi leptonic and hadronic decays of the 
top quarks respectively. Note that the final states consist multiple 
$b$ quarks, a characteristics of the heavier charged Higgs signal for the
{\Dchtb}
decay channel~\cite{Miller:1999bm,Moretti:1999bw}.  
The top quark originating from $\chh$ decay is tagged 
in it's hadronic mode, and combined with the appropriately identified $b$-jet, the charged Higgs mass
reconstructed. Tagging of top quark is performed implementing the powerful jet substructure 
analysis\cite{Butterworth:2008iy}, which is postponed for discussion in the
next section. In case of pure hadronic signal final state, the associated top 
quark is also identified through kinematic reconstruction in order 
to make signal more robust. In addition to the 
reconstruction of top quarks, we exploit the presence of extra   
hard b-jets in the final state in order to separate out backgrounds. 
Therefore, we focus the charged Higgs signal final state in two categories:
\begin{equation}
\begin{array}{ccccccc}
\text{a} & : & {H}_{\text{reco}}^{\pm} & + & {t}_{\text{reco}} & + & \text{n}_{ b}\left(\ge 1\right)~{ b}-\text{jet}\\
\text{b} & : & {H}_{\text{reco}}^{\pm} & + & \text{n}_{\ell}\left(\ge1\right) & + & \text{n}_{ b}\left(\ge1\right)~{ b}-\text{jet}
\end{array}\label{eq:sig2}
\end{equation}
where $\chh_{\text{reco}}$ and ${ t}_{\text{reco}}$ represent the reconstructed
Charged Higgs and top quark,
and $n_{\ell}$ and $n_{ b}$ are the number of leptons and $b$-jets
respectively and required to be at least one.
The main dominant source of irreducible SM backgrounds are due 
to $\ttbar$,
and inclusive hard QCD jet production.
However, in both cases, extra b-jets may arise via gluon splitting in the initial state radiation.
The QCD jet production becomes dominant source of irreducible background, 
in particular corresponding to the hadronic signal final state, 
due to the non-negligible mis-tagging probability of hard jets 
as a top jet. 
Moreover, the process $\ttbar { g}$ which predominantly 
produces the final state $\ttbarbbbar$ is also taken into account in our
background estimation. Before discussing the signal and background 
simulation strategy,   
we discuss briefly the top tagging methodology used in our simulation.

\subsection{Top Tagging}
\label{toptagging}
\begin{figure}[H]
\centering
\includegraphics[width=0.6\textwidth]{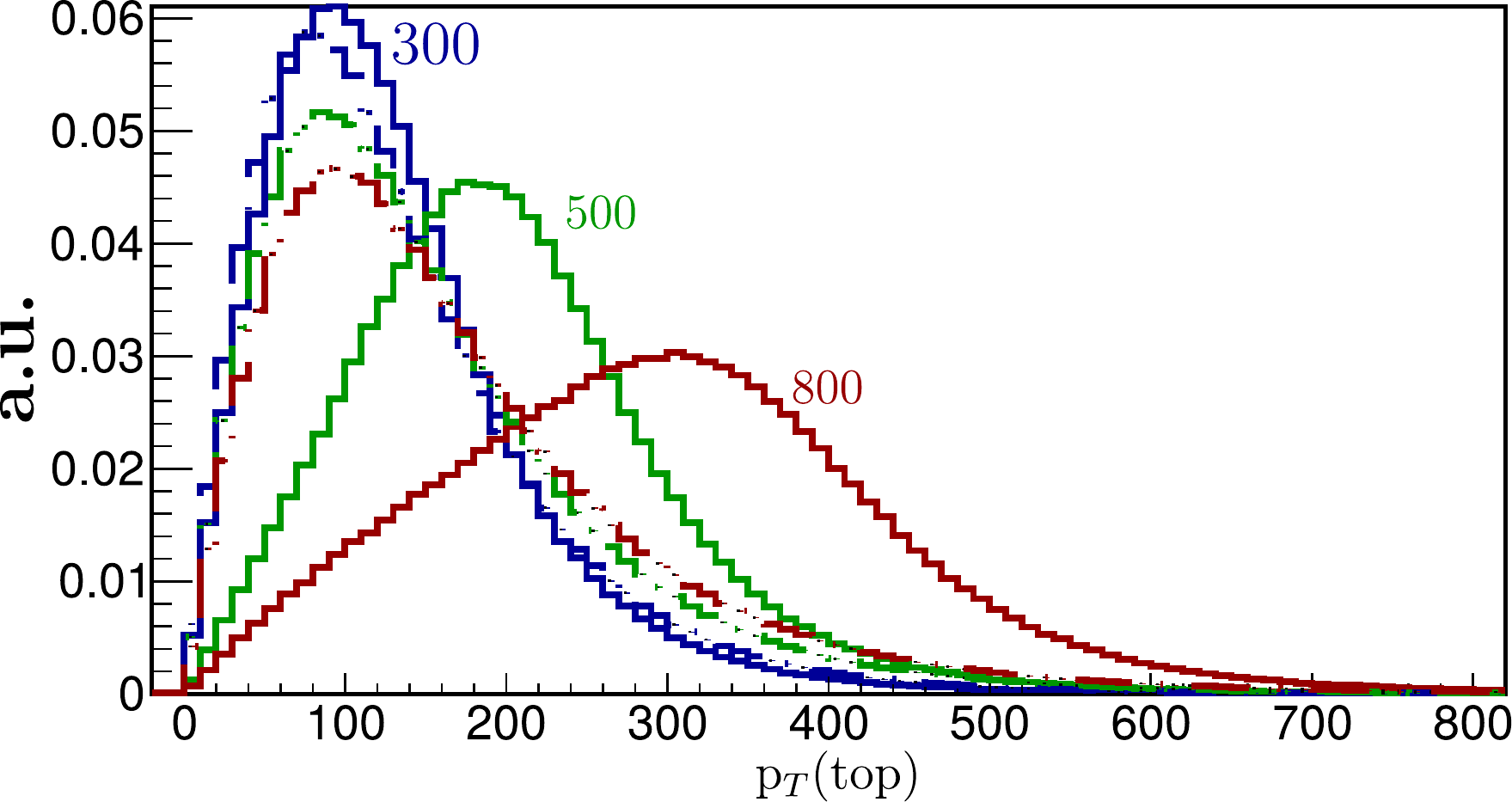}
\caption{
	Parton level transverse momentum of top quark from  
	charged Higgs decay(solid) and associated with it(dashed) 
	for $\mch=300$ GeV, 500 GeV 
	and 800 GeV, normalized to arbitrary units (a.u).
	The distribution of $p_T$ of top quark corresponds to the combined processes (Eq.~\ref{eq:Production}) and then appropriately matched using Eq.~\ref{eq:4FS}.
}
\label{fig:pttop}
\end{figure}
The top tagged jets are the essential
components of our considered signal events.  
It has been pointed out earlier that the top quark originating
from $\chh$ decay is expected to be
boosted(boost factor, $\gamma_{ t} \sim \mch/m_{ t}$),
in particular for heavier charged Higgs boson masses. 
The $p_T$ of those top quarks are demonstrated
in Fig.~\ref{fig:pttop} 
for three masses of $\chh$, along with the same for associated top quarks.
Clearly, this figure indicates that the top quark from heavier $\chh$ decay 
is moderately boosted, however, $p_T$ distribution of associated top quarks
are found not so sensitive to $\mch$.
A top quark decays to a $b$-quark and a $W$
which subsequently decays to a pair of light quarks leading to jets in 
the calorimeter.
However, for fast top quark, these decay products may not
appear well separated to resolve as a separate jets. In such cases, the
boosted top quarks may look like a single jet, called fat jet with three
or more subjets as constituent corresponding to its decay products.
These subjets are well separated within an angular cone of the
order $\sim 2m_{ t}/p_T$. Following this kinematic features, we attempt to 
tag topjets, surrounded by busy hadronic environment using the 
top tagger, namely HepTopTagger~\cite{Plehn:2011tg,Kasieczka:2015jma,Plehn:2010st,Plehn:2009rk}.
In the process of tagging tops,  
first cluster particles with 
$p_T\ge 0.5$~GeV and $|\eta|<5$, using the Cambridge/Aachen \cite{Dokshitzer:1997in} jet
algorithm implemented in {\tt Fastjet-3.3.0} \cite{Cacciari2012}
for jet radius R=1.5 to form fat jets. Then require at least one
hard fat jet in the event with $p_T\ge 200$~GeV. In our searches, 
top tagged jets are likely to be contaminated by QCD radiation, since
wider radius R=1.5 is considered to contain all subjets from the
moderately boosted top quark decay. Therefore, it is suggestive 
to take extra measures to eliminate QCD effects due to soft radiation 
in reconstructing subjets.
The sub structures of Fatjets are obtained following the mass drop method 
using some recursive steps which are built in 
HepTopTagger\cite{Plehn:2011tg,Kasieczka:2015jma,Plehn:2010st,Plehn:2009rk}. 
In this process, the last step of clustering process
is declustered to obtain two subjets $j_1$ and $j_2$, such that
$m_{j_1} > m_{j_2}$. If $m_{j_1} + m_{j_2} \sim m_j$,
and $m_{j_1} >0.8 m_j$, then it is expected that $j_2$ originates
from QCD emission or underlying events, and we discard $j_2$, otherwise
we keep both $j_1$ and $j_2$.
If the mass of the subjet is 30 GeV or less, then we keep it or 
decompose it further (both $j_1$ and $j_2$ or just $j_1$ depending on how
symmetrically the mass splits). The subjets which are obtained at the end
of the recursive declustering procedure, are also cleaned further through
filtering~\cite{Butterworth:2008iy} to eliminate the contamination
from the QCD radiation. Two subjets are suppose
to originate from $W$ decay, and it is
ensured by requiring the invariant mass of two subjets
$m_{jj} = m_{W} \pm 15$~GeV. Finally, the top is tagged by adding the
third sub-jet, which is a $b$-like jet, examined
by matching with the parton level b quark in the event. The invariant mass of three
subjets after filtering is required to be $m_{jj{ b}} = m_{ t} \pm 30$~GeV.
If there be more than one top tagged jet, we choose the one which is 
the closest to the pole mass of the top quark. Using the default 
conditions in HepTopTagger, we find the single top tagging efficiency is 
about 10\% for this kind of moderately boosted tops in signal events. 
Note that in calculating this efficiency no
pile-up effects are taken into account. The mistagging efficiencies are
obtained using the QCD events and it is found to be around $2-3$ \%.

\begin{figure}[H]
	\caption{ROC curve for the MVA TopTagger obtained 
		from signal events in hadronic final state and QCD for mis tagging. 
		Efficiencies obtained  using HepTopTagger is also shown.}
	\centering
	\includegraphics[width=0.7\linewidth]{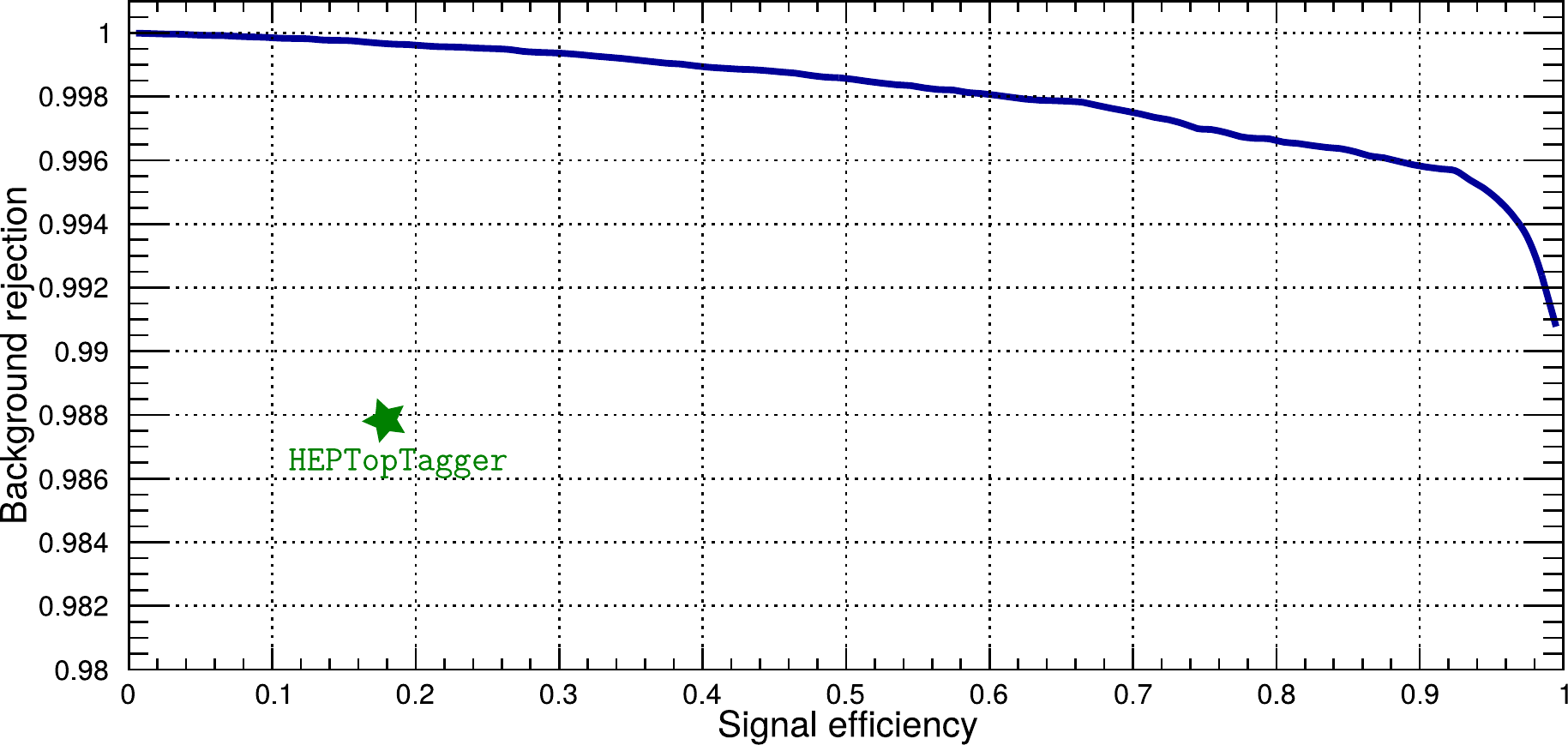}
	\label{fig:mvatoptagger}
\end{figure}

We attempt to recover this top tagging efficiency to a better level
by employing multivariate analysis.
The multivariate analysis is implemented within TMVA~\cite{Hocker:2007ht}
combining the HepTopTagger mass drop method, and instead of using the
full chain of HepTopTagger, some other additional
kinematic variables including N-Subjettiness, energy correlation,
are used as listed below:
\begin{enumerate}
\item{N-Subjettiness}~\cite{Thaler:2011gf}:
Variables are defined as,
\br
\frac{\tau_2}{\tau_1}, \frac{\tau_3}{\tau_2}, \frac{\tau_4}{\tau_3},
\er
where $\tau_N$ is the N-th subjettiness
variable~\cite{Thaler:2010tr} as defined,
\br
\tau_N = \frac{1}{R_0 \sum_k P_{T,k} } \sum_k P_{T,k} {\rm min}
\left (\Delta R_{1,k}, \Delta R_{2,k}...\Delta R_{N,k} \right)
\er
$\Delta R_{ik}$ is defined to be the geometrical separation
between i-the subjet and the k-th reference axes, $R_0$ is the jet
cone size parameter. Clearly, a smaller $\tau_N$ implies more radiation
around the given axes, i.e a better description of jets with N or
less subjets, where as large $\tau_N$ means a better description of
jets with more than N subjets. It is found that $\tau_N/\tau_{N-1}$ is
an efficient discriminating variable to distinguish boosted
objects \cite{Thaler:2010tr,Stewart:2010tn,Thaler:2011gf}.
\item{Mass difference:}
It is defined as, 
$\Delta m_t = |m_{j_t} - m_{t}|$, where $m_{j_{t}}$ is the mass of
the tagged top jet. This mass difference is also very crucial in
tagging tops.                                          
\item{Invariant mass of 2 and 3 subjets:}
Invariant mass of the 3 sub jets, $m_{123}$ and 2 sub jets $m_{ij}$  where
$(i,j\in\left\{1,2,3\right\})$, is computed for each possible
combination of subjets.
\item{Number of $b$-like sub jets}:
The number of $b$-like sub jets $n_{b}^j$, it is counted by matching
subjets with $b$-partons within $ |\eta| < $ 2.5, and $p_T> 5$~GeV
using the matching cone $\Delta R<0.3$ around the subjet.
\item{Variable related with reconstructed masses:}
It is defined to be,
\br
f_{\text{rec}}\equiv\min_{i,j}\left|\frac{\left(\frac{m_{ij}}{m_{123}}
\right)}{\left(\frac{m_{W}}{m_{t}}\right)}-1\right|.
\er
This ratio determines the quality of reconstructed $W$ with respect
to the overall quality of reconstructed top mass.
\item{Energy correlations:} The energy correlators among the subjets
or particles inside a jet distinguishes the various properties
of jets \cite{Larkoski:2013eya}.
The correlation function uses the information about the energies and
pair-wise angles between particles within a jet. This generalized energy
correlation function, is interestingly can be made infrared and collinear 
safe. This energy correlation function is found to be very 
effective to classify jets. For details, see Ref.\cite{Larkoski:2013eya}.
\end{enumerate}
With these set of variables, 1-6,      
we train Boosted Decision Trees to
tag top jets in $\ttbar$ and mis-tags in QCD process.
In Fig.\ref{fig:mvatoptagger}, we show the results as a 
receiver operator response(ROC) curves
for both signal acceptance and background(QCD) rejection efficiencies.
This figure clearly demonstrates an improvement in top tagging efficiencies,
along with the suppressed background mis-tag rates.
The efficiencies obtained using HepTopTagger  
is also shown by a star.
Undoubtedly, the top tagging efficiency through MVA method 
is improved significantly. We use this improved efficiency 
in the simulation of signal and background.

\subsection{Signal and Background Simulation}
\label{simulation}
The {\pythia}~\cite{Sjostrand:2006za} is used to generate events via 
the process {\signalgbth}, where as {\madgraph}~\cite{Alwall:2014hca} is used
for {\signalggtbh} and then showering through {\pyth}.
The dominant SM background processes {\ttbar} and QCD events are 
generated using {\pyth}, while {\mg} interfacing 
with {\pyth} is used for {\ttbarbbbar} process.
Events are generated by dividing the phase space in $\hat p_T$ bins, 
$\hat p_T$ is the transverse momentum of the final state partons in the
center of mass frame. 
For instance, in case of signal events, bins are chosen 
as $0-200$ GeV, $200-400$ GeV and $400-\infty$, where as for backgrounds ({\ttbar} and QCD),
bins are set as $0-200$ ($20-200$ for QCD) GeV, $200-500$ GeV, $500-800$ GeV and $800-\infty$.
Various event selections imposed in the  simulation 
for both signal and backgrounds are described below:
\begin{enumerate}
\item{Lepton selection:}
Leptons, both electrons and muons are selected with cuts on 
the transverse momentum($p_T^\ell$), and rapidity($\eta_\ell$),
\br 
p_T^{\ell}\ge20~{\rm GeV},~~~~~\left|\eta_{\ell}\right|\le2.5
\er
Isolation of lepton is ensured by requiring, $E_T^{AC} \le$ 30\% of 
$p_T^\ell$, where $E_T^{AC}$ is the sum of transverse momenta of the particles 
which are within the cone $\Delta R$(= $\sqrt{\Delta \eta^2 + \Delta \phi^2}$)
$<0.3$ along the direction of lepton.
It is to be noted that the lepton isolation criteria is not imposed  
while selecting events applying lepton veto, otherwise genuine leptonic
events would contribute to the hadronic events.

\item{$b$-jet identification:}
{In the simulation jets are reconstructed using 
Fastjet~\cite{Cacciari2012} 
with anti-$k_T$ algorithm \cite{Cacciari:2008gp}
and jet size parameter R=0.5. Reconstructed jets are subject to
$p_T^j>20$GeV, $|\eta_j|<4$.
A given reconstructed jet  
identified as $b$-like jet, if there is a matching with parton 
level $b$ quark with a matching cone
$\Delta R < 0.3$. In addition, the matched jets are required to have
$|\eta|<2.5$.     
We found about 70\% cases $b$ quarks are identified as $b$-like jets 
Finally, in the simulation to select b-like jets, 
we apply a hard cut $p_T>$30~GeV. It is to be noted that in our simulation
the mistags are not taken into account, which is out of the scope of the
present analysis. However,from the studies \cite{ATL-PHYS-PUB-2015-022,CMS-PAS-BTV-15-001}, we found that the
mistags of the order of few percent are not expected to affect our results 
significantly.}      
\item{Top reconstruction:} 
The details of the top tagging are already discussed in the previous section.
{However among the tops tagged using this technique, 
we found 60-70\% are from
the decay of {\chh} while the remaining are the associated tops,
for the case of $\mch=500$ GeV and it goes up with the increase of $\mch$.}
In addition, after the reconstruction of 
charged Higgs using tagged top jets, an additional top quark is also 
reconstructed through kinematic fitting out of the remaining jets 
for hadronic signal events.
This extra kinematically reconstructed top quark is likely to 
correspond to the associated top quark.
For leptonic signal events no such top quark is reconstructed.

\item{Charged Higgs mass reconstruction:}
We observed via matching that the leading identified $b$-jet with 
$p_T>50$~GeV corresponds ($\sim$ 70-80\%) to $b$ quark 
originating from $\chh$ decay (for ${\mch} {\gtrsim} 500$ GeV). Hence, 
the charged Higgs mass is reconstructed combining the leading top 
tagged jet with the leading $b$-like jet. 
In Fig.\ref{fig:CHreco}, we show the reconstructed 
mass($m_{{t}_j {b}_1}$) of charged Higgs for three input 
values, $\mch=300$~GeV, 500~GeV, 800~GeV along with the dominant 
background from $t\bar t$ corresponding to hadronic final states,
Eq.~\ref{eq:sig2}(a) subject to selection cuts on $b$-jets.
The distribution due to $t\bar t b \bar b$ appears to be almost same 
as $t\bar t$, where as for QCD it comes out as flat without any visible peak.
The distributions from both these sources are not shown in this 
Fig.~\ref{fig:CHreco}, otherwise it would be very crowded.  
Notice that the peaks are not appearing exactly at the 
input mass of charged Higgs because of the smearing of the momenta of 
tagged top and $b$ jet.
The wide spread of $m_{{t}_j {b}_1}$ distribution around the 
peak is due to incorrect combination of the reconstructed top  
and $b$-like jet.
The events are selected 
requiring the reconstructed mass $m_{{t}_j {b}_1}$ within the range, 
\br
m_{{t}_j {b}_1}=m_{\chh} \pm 0.3~m_{\chh},
\label{eq:MHreco}  
\er
which is 30\% around the peak. 
\begin{figure}[h]
        \begin{center}
        	\caption{Charged Higgs mass($=m_{{{t}_j}{b}_1}$, 
matched with 4FS and 5FS) reconstruction for 
$\mch=300$~GeV, 500~GeV and 800~GeV and $\tan\beta=30$ along with the 
background from $t \bar t$(dashed).}
            \includegraphics[scale=0.7]{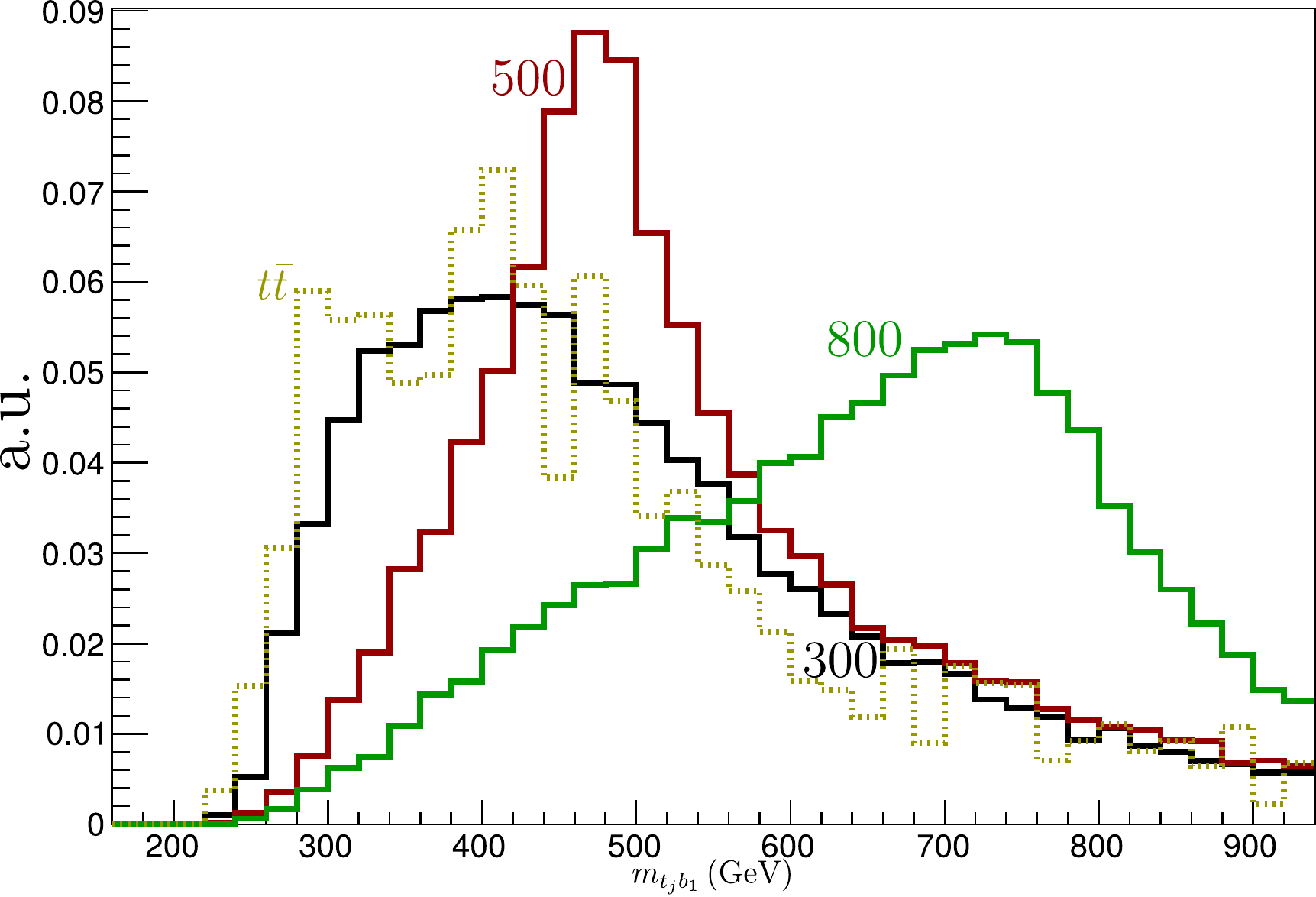}
            \label{fig:CHreco}
        \end{center}
\end{figure}
 
\item
{Multiplicity of $b$ jets:} 
In signal events multiplicity of $b$-jets is higher than the 
$\ttbar$ and QCD backgrounds. Hard $b$-jet remains in signal final state,
even after reconstruction of two(one) tops, 
and subsequently a charged Higgs in hadronic(leptonic) final state.  
The additional $b$-jets appearing in  
background events are due to the gluon splitting, and is not 
expected to be hard. 
Therefore, requirement of at least one hard $b$ jet in the final state 
is expected to be useful in rejecting backgrounds. 
Hence, a selection,  
\br
n_{b} \ge 1 \ \ {\rm with}\ \  p_T^{b} \ge 30~{\rm GeV} 
\er
is imposed in the simulation.
\end{enumerate}

\subsection{Results}
\label{cutbasedresults}
We simulate both the production processes
in 4FS and 5FS, and then obtain the final yield
by appropriately weighting both the resulting cross sections,
as per prescription given by Eq.~\ref{eq:4FS} and \ref{eq:5FS}.
For the illustration purpose, in Table~\ref{tab:THad}, the event yields 
in terms of cross sections are presented after each set of cuts
as described above, for signal
and backgrounds corresponding to the hadronic final state, 
see Eq.~\ref{eq:sig2}(a). The second row shows the 
total production cross sections of the respective processes
at 13~TeV center of mass energy. 
Results for signal events are shown only for a representative choice of
a single mass of charged Higgs, $\mch=500$~GeV, although simulations are 
performed for a wide range of masses, upto 1 TeV.
Also note that the results are presented 
for $\tan\beta=30$ and within the framework of Supersymmetric based model
(Type II).
Fat jets are reconstructed
selecting events with a lepton veto and at least one 
$b$-identified jets.
In order to access the boosted region, events are selected
with a $p_T>200$~GeV on fat jets. These high $p_T$ 
fat jets are used as a input to HepTopTagger to tag them as top jets.
We employ MVA method as described above to tag top jets, and  
found that about 30\% of events are tagged as a top jet.
Subsequently, after top tagging, we look for the hardest leading 
$b$-jet with a cut $p_T>50$ GeV, which is found to be 
originating from $\chh$ decay for about 70-80\% events.
Combining top tagged jets and the hardest $b$ jet, the charged Higgs mass is
reconstructed, and select events within
the mass window $\pm 30\%$ of the input charged Higgs mass.
Notice that a good fraction of background events remain within 
this reconstructed charged Higgs mass window. With the remaining 
untagged jets and identified $b$-jets, the associated top quark is 
reconstructed. The requirement of a second reconstructed top quark, 
suppresses the background, in particular QCD,
more than the signal. Finally, demanding a hard $b$-jet with $p_T>30$~GeV
rejects backgrounds substantially.
\begin{table}[H]
	\begin{center}
		\caption{
			Cross section yields for signal and backgrounds in the 
			hadronic signal final state (Eq~\ref{eq:sig2}(a)) for ${\mch}=500$~GeV, $\tan\beta=30$ in Type II 2HDM. The first row presents the production cross sections. For signal, 
			{\Brhtb} is multiplied with the signal cross sections.
		}
		\label{tab:THad}
		\begin{tabular}{|c|ccccc|}
			\hline
			Selection  & 5FS$\times$Br & 4FS$\times$Br & $\ttbar$  & $\ttbarbbbar$  & QCD  \\
			\hline
			$\sigma$(fb) & {$124.9$}  & {$64.4$}  & {$8.3 \times 10^{5}$}  & {$1.4 \times 10^{4}$}  & {$7.2 \times 10^{11}$}  \\
			
			$N_{ b}\ge1$ \& Lepton Veto  & {$55.8$}  & {$28.8$}  & {$4.6 \times 10^{5}$}  & {$7.4 \times 10^{3}$}  & {$1.2 \times 10^{10}$}  \\
			
			$N_{\text{FJ}}\ge1$  & {$44.8$}  & {$24.2$}  & {$1.1 \times 10^{5}$}  & {$2.6 \times 10^{3}$}  & {$9.4 \times 10^{6}$}  \\
			
			$N_{{ t}_j} \ge 1 $  & {$13.3$}  & {$7.4$}  & {$3.2 \times 10^{4}$}  & {$790.4$}  & {$2.5 \times 10^{5}$}  \\
			
			$p_{T}^{{b}_{1}}\ge50$ GeV  & {$12.3$}  & {$6.9$}  & {$2.0 \times 10^{4}$}  & {$633.5$}  & {$9.2 \times 10^{4}$}  \\ 
			
			$m_{{ t}_{j}{ b}_{1}}\in\left[0.7 \mch ,1.3 \mch \right]$  & {$8.8$}  & {$5.2$}  & {$1.3 \times 10^{4}$}  & {$387.6$}  & {$4.2 \times 10^{4}$}  \\
			
			$N_{{ t}_{\text{Associated}}^{\text{Hadronic}}}=1$  & {$2.2$}  & {$1.6$}  & {$364.1$}  & {$78.8$}  & {$1.2 \times 10^{3}$}  \\
			
			Extra b, $p_T \ge 30$ GeV  & {$0.5$}  & {$0.5$}  & {$20.3$}  & {$15.9$}  & {$50.2$}  \\
			\hline
		\end{tabular}
	\end{center}
\end{table}
Similarly cross section yields for leptonic final
states(Eq.\ref{eq:sig2}(b)) are presented in Table~\ref{tab:lepcs}.
The events are selected with at least one
identified $b$-jet and one isolated lepton.
A top jet is tagged, and it is observed that
efficiency of top tagging is less, due to the lack of
availability of many hadronic top quarks.
As before, requiring a hard identified $b$-jet, with
$p_T>50$ GeV and combining it with tagged top jet, the charged Higgs mass is
reconstructed. Finally, requirement of a hard $b$-jet suppresses the background
more than the signal. Use of an
additional cut on missing transverse
momentum due to the presence of neutrinos in the leptonic decay of top quark
is found to be not so helpful.
\begin{table}[H]
	\begin{center}
		\caption{Same as in Table~\ref{tab:THad},  but for the leptonic signal final state.}
		\label{tab:lepcs}
		\begin{tabular}{|c|ccccc|}
			\hline
		Selection  & 5FS$\times \text{Br}$ & 4FS$\times \text{Br}$ & $\ttbar$  & $\ttbarbbbar$  & QCD  \\
			\hline
			$\sigma$(fb)& {$124.9$}  & {$64.4$}  & {$8.3 \times 10^{5}$}  & {$1.4 \times 10^{4}$}  & {$7.2 \times 10^{11}$}  \\
			
			$N_{ b}\ge1$ \& $N_{\ell}\ge1$  & {$39.3$}  & {$20.3$}  & {$2.5 \times 10^{5}$}  & {$4.2 \times 10^{3}$}  & {$5.0 \times 10^{6}$}  \\
			
			$N_{\text{FJ}}\ge1$  & {$29.3$}  & {$15.8$}  & {$5.0 \times 10^{4}$}  & {$1.3 \times 10^{3}$}  & {$3.6 \times 10^{3}$}  \\
			
			$N_{{t}_j}\ge1$  & {$5.4$}  & {$3.0$}  & {$1.0 \times 10^{4}$}  & {$276.1$}  & {$103.1$}  \\
			
			$p_{T}^{{b}_{1}}\ge50$ GeV  & {$4.9$}  & {$2.8$}  & {$6.7 \times 10^{3}$}  & {$221.8$}  & {$71.7$}  \\
			
			$m_{{ t}_{j} { b}_{1}}\in\left[0.7 \mch,1.3 \mch\right]$  & {$3.4$}  & {$2.0$}  & {$4.4 \times 10^{3}$}  & {$138.3$}  & {$20.5$}  \\
			
			$p_{T}^{{b}_{2}}\ge30$ GeV  & {$2.1$}  & {$1.3$}  & {$301.0$}  & {$66.1$}  & {$\lesssim 1.0$}  \\
			
		 \hline
		\end{tabular}
	\end{center}
\end{table}

\begin{table}[H]
	\begin{center}
		\caption{
			Cross sections (fb) normalized by acceptance efficiencies ($\epsilon_{\text{ac}}$)
			for signal and background. Signal significances for three integrated
			luminosity options for hadronic (leptonic) final state are performed for the
			Type II model and $\tan \beta=30$.
		}
		\label{tab:HadSbyB}
		\begin{tabular}{|c|ccc|}
			\hline 
			&  & $\sigma\times\epsilon_{\text{ac}}$ (in fb)  & \tabularnewline
			${\mch}{\rightarrow}$ & 300 GeV  & 500 GeV  & 800 GeV \tabularnewline
			\hline 
			5FS  & 0.4 $\left(1.4\right)$ & 0.5 $\left(2.1\right)$ & 0.1 $\left(0.51\right)$\tabularnewline
			4FS  & 0.3 $\left(0.95\right)$ & 0.5 $\left(1.3\right)$ & 0.1 $\left(0.34\right)$\tabularnewline
			$\ttbar$  & 5.9 $\left(140.0\right)$ & 20.3 $\left(301.0\right)$ & 15.5 $\left(142.3\right)$\tabularnewline
			$\ttbarbbbar$  & 5.4 $\left(22.0\right)$ & 15.9 $\left(66.1\right)$ & 8.8 $\left(37.6\right)$\tabularnewline
			QCD  & $\lesssim 1.0$ $\left(\lesssim 1.0\right)$ & 50.2 $\left(\lesssim 1.0\right)$ & 21.4 $\left(\lesssim 1.0\right)$\tabularnewline
			\hline 
			Matched Signal cross section(S)  & 0.4 $\left(1.3\right)$ & 0.5 $\left(1.9\right)$ & 0.1 $\left(0.47\right)$\tabularnewline
			Total Background cross section(B)  & 11.3 $\left(161.9\right)$ & 86.4 $\left(367.1\right)$ & 45.7 $\left(179.9\right)$\tabularnewline
			\hline 
			${\cal L}$~(fb$^{-1}$)  &  & $S/\sqrt{B}$  & \tabularnewline
			300  & 1.9 $\left(1.73\right)$ & 0.92 $\left(1.71\right)$ & 0.28 $\left(0.61\right)$\tabularnewline
			1000  & 3.4 $\left(3.16\right)$ & 1.7 $\left(3.13\right)$ & 0.51 $\left(1.11\right)$\tabularnewline
			3000  & 5.9 $\left(5.48\right)$ & 2.9 $\left(5.42\right)$ & 0.88 $\left(1.92\right)$\tabularnewline
			\hline 
		\end{tabular}
	\end{center}
\end{table}
In Table~\ref{tab:HadSbyB}, we summarize the 
signal and background cross sections normalized by the 
kinematic acceptance efficiencies for both the hadronic and leptonic 
final state respectively.
For illustration, we show results for three choices of charged 
Higgs mass, $m_{\chh}=300$, 500 and 800 GeV, corresponding to 
the signal cross sections in both 4FS and 5FS mechanisms.
The signal cross sections are found to be 
${\cal O}$(fb), where as the total background contribution is huge, in
particular for hadronic final state. But for leptonic final state,
the level of background contamination is comparatively less. In this case, the presence of  
leptons and a hard $b$ jet requirement in the final state
help to get rid of a good fraction of the QCD background.

The signal significances are presented for three integrated luminosity 
options ${\cal L}=300$, 1000 and 3000 $\invfb$. 
Table \ref{tab:HadSbyB} reveals that the charged Higgs boson of mass 300 GeV
can be discovered for high luminosity options(3000 $\invfb$)
with a reasonable significance, but for higher masses
$\sim 500$~GeV or more, the signal is  merely observable. Clearly, 
it is hard to achieve discoverable signal sensitivity 
for heavier charged Higgs mass in this channel. 
However, discovery potential of charged Higgs in leptonic 
final state is comparatively better. 
For instance, Table~\ref{tab:HadSbyB} shows that the charged Higgs signal observable
with a moderate significance for the mass range around 500 GeV
even for 1000 $\invfb$ integrated luminosity option.

In summary, undoubtedly, this cut based analysis indicates how 
difficult it is to achieve discoverable sensitivity of charged Higgs 
signal in the ${t}\bar{ b}$ decay mode owing to the huge background cross
section with identical event topology. The present set of     
cuts are not very efficient to suppress backgrounds at the required level
in order to make signal sensitivity better.  One may  
think of more better construction of kinematic observables, and 
devise a set of cuts providing efficient optimization to reduce 
the background effect. It is a very challenging task to find the feasibility of the charged Higgs signal 
for heavier masses at the LHC.   
It motivates us further to develop a search strategy 
using the technique of multivariate analysis, 
which is discussed in the next section.

\section{Multivariate Analysis}
\label{MVAResults}
In the previous section, we observed that there is
no single or a combination of kinematic variables which has the 
potential to isolate tiny signal out of huge backgrounds. 
In this section, we discuss MVA in order 
to improve signal to background ratio aiming to achieve a 
better significance for a given luminosity option.
The basic idea of this method is to combine many kinematic variables
which are the characteristics of signal events, into a single
discriminator, and eventually this single discriminator is used
to separate out the signal suppressing backgrounds.
The MVA framework is a powerful tool used very widely in high energy 
physics, to extract the tiny signal events out of huge background 
events, including single top discovery~\cite{Abazov:2006gd} and 
recently the Higgs boson at the LHC~\cite{Khachatryan:2014ira}.
Here we carry out MVA through Boosted Decision Tree(BDT) method
within the framework of TMVA \cite{Hocker:2007ht}. 
 
In the BDT method, events are classified by applying sequentially a set of 
cuts making sub sets of events with different signal purity. 
Several disjoint decision trees consisting two branches are constructed 
through a best selection of cuts out of listed input variables of the 
given process, and it is repeated using subsequent set of cuts till 
all the events are classified. 
While training the sample 
events, if an event is misclassified i.e a signal event labeled as 
background or background event as signal event, then it is boosted by 
increasing the weight of that event. Subsequently, a 
second tree is made using the new weights, which may not be same 
as the previous tree. This process is repeated
and we constructed about 1000 trees. There are few methods of 
boosting \cite{Schapire1990}, and we use the 
gradient boosting technique \cite{friedman2001}.
In BDT algorithm, these trees are made by training half of the 
signal and background events. The remaining half of the 
signal and background events 
are used to check the performance of the trained BDT.

Following the production and decay mechanism, Eq.~\ref{eq:sig},
events are selected for the final state consisting 
one top tagged jet, more than one identified $b$ jet and 
untagged jets corresponding to hadronic signal final state.
For leptonic signal, in addition, at least one isolated lepton
is required.  
A large number of kinematic variables are constructed out of 
the momenta of these 
objects to train event samples, and eventually 10 input variables 
are used in BDT to train signal 
and background sample.  
In Table \ref{tab:hadVar} and \ref{table:varlep},
the set of input variables are shown ranking them according to the importance 
in the BDT analysis for ${\mch}=500$~GeV corresponding to hadronic and 
leptonic final states respectively. In the third row of this table
a brief description is provided for each of the variables.  
The importance here means the effectiveness of those variables in 
suppressing backgrounds while maintaining a better signal purity. 
\begin{table}
	\begin{center}
		\caption{Kinematic variables used to train the signal and 
			background sample for hadronic final state and $\mch=500$ GeV.}
		\label{tab:hadVar}
		\begin{tabular}{|ccc|}
			\hline
			Rank & Variables & Description \\
			\hline
			1 & $m_{{ b}\bar{ b}}^{12}$ & Invariant mass of two b jets \\
			2 & $p_{T}^j$ & $p_T$ of the leading jet \\
			3 & TaggedTopMVA & MVA Discriminator for Toptagging \\
			4 & $m_{{ b}_1 { b}_2 { b}_3}$ & Invariant mass of 3 bjets \\
			5 & $p_{T_2}^{{ b}_{\text{jet}}}$ & $p_T$ of 2nd b jet after top tagging \\
			6 & $m^{\text{reco}}_{{ t}_j {b}_1}$ & Reconstructed Higgs mass \\
			7 & $H_T$ & Scalar sum of $p_T$ of all final detectable particles\\
			8 & $n_j$ & Number of un tagged \\
			9 & $p_{T_3}^{{ b}_{\text{jet}}}$ & $p_T$ of 3rd b jet \\
			10 & $H_T^{ b}$ & Scalar sum of $p_T$ of all b jets\\
			\hline   
		\end{tabular}
	\end{center}
\end{table}

\begin{table}
	\begin{center}
		\caption{Same as Table~\ref{tab:hadVar}, but for leptonic final state.}
        \label{table:varlep}
		\begin{tabular}{|ccc|}
			\hline
			Rank & Variables & Description \\
			\hline
			1 & $H_T^{ b}$ & Scalar sum of $p_T$ of all b jets\\
			2 & $m_{{ t}_j}$ & Mass of the tagged top jet \\
			3 & $m_{{ b}\bar{ b}}^{12}$ & Invariant mass of two b jets \\
			4 & $m_{{ t}{{ b}_2}}$ & Invariant mass of tagged top jet and second b jet \\
			5 & $H_T/{\text{MHT}}$ & Ratio over $H_T$ and MHT \\
			6 & $p_T^{j_1}$ & $p_T$ of leading un tagged jet \\
			7 & $H_T$ & Scalar sum of $p_T$ of all jets \\ 
			8 & $p_{T}^{{ b}_{\text{jet}}}$ & $p_T$ of leading b jet \\
			9 & $p_{T_2}^{{ b}_{\text{jet}}}$ & $p_T$ of $2^{\text{nd}}$ b jet after top tagging\\
			10 & MHT & Vector sum of $p_T$ of all jets and leptons \\
			\hline
		\end{tabular}
	\end{center}
\end{table}
We have observed that for ${\mch}=300$ GeV, the importance or ranking 
of some of the variables are altered. For example, 
for 300 GeV, $p_{T_3}^{{b}_{\text{jet}}}$ and $H_T^{b}$ are found to be more important 
than $m^{\text{reco}}_{tb}$. Similarly, for very heavier charged Higgs mass, 
the $H_T$ is expected to be more important in suppressing background, 
hence it is ranked to second. Interestingly, the invariant mass of the first
two leading $b$ jets seems to be a very strong discriminant variable in 
separating the signal and background. 
Moreover, the MVA discriminator for top tagging using 
HepTopTagger, multiplicity of untagged jets, 
and $p_T$ of the second $b$-jet, all appear to be useful variables
in eliminating the background events. 

In Table~\ref{table:varlep}, the set of kinematic variables are presented for
leptonic final state and for ${\mch}=500$~GeV. However, as before, this set 
remains 
same for ${\mch}=300$ and 1000~GeV, but ranking becomes different for 
obvious reasons. For instance,
for lower mass of ${\mch}=500$~GeV, the variable 3 becomes more important
than the variable 1. Due to the presence of neutrinos, the variable
related with missing transverse energy, MHT plays role in discriminating
background, in particular from QCD. Like hadronic case, the number 
of $b$ jets and their corresponding transverse  momentum 
are very effective in increasing signal to background ratio. 

In this type of analysis based on machine learning models, one of the issue
often encountered is the problem of overtraining the sample. The training of 
the sample can be checked using a test data sample. Ideally, for a 
sufficiently large and random monte carlo data, performance of 
training and testing data should be similar. If a significant deviations 
between these two are found, that 
would be an indication of over training of the sample. 
This overtraining tests are performed for all ${\mch}=300-1000$~GeV 
masses.  
\begin{figure}[H]
    \begin{center}
        \caption{MVA output (D) distribution for signal and backgrounds corresponding to hadronic signal final state and ${\mch}=500$ GeV, $\tan \beta = 30$ for Type II 2HDM.}
        \label{fig:Ddist}
        \includegraphics[width=0.8\textwidth]{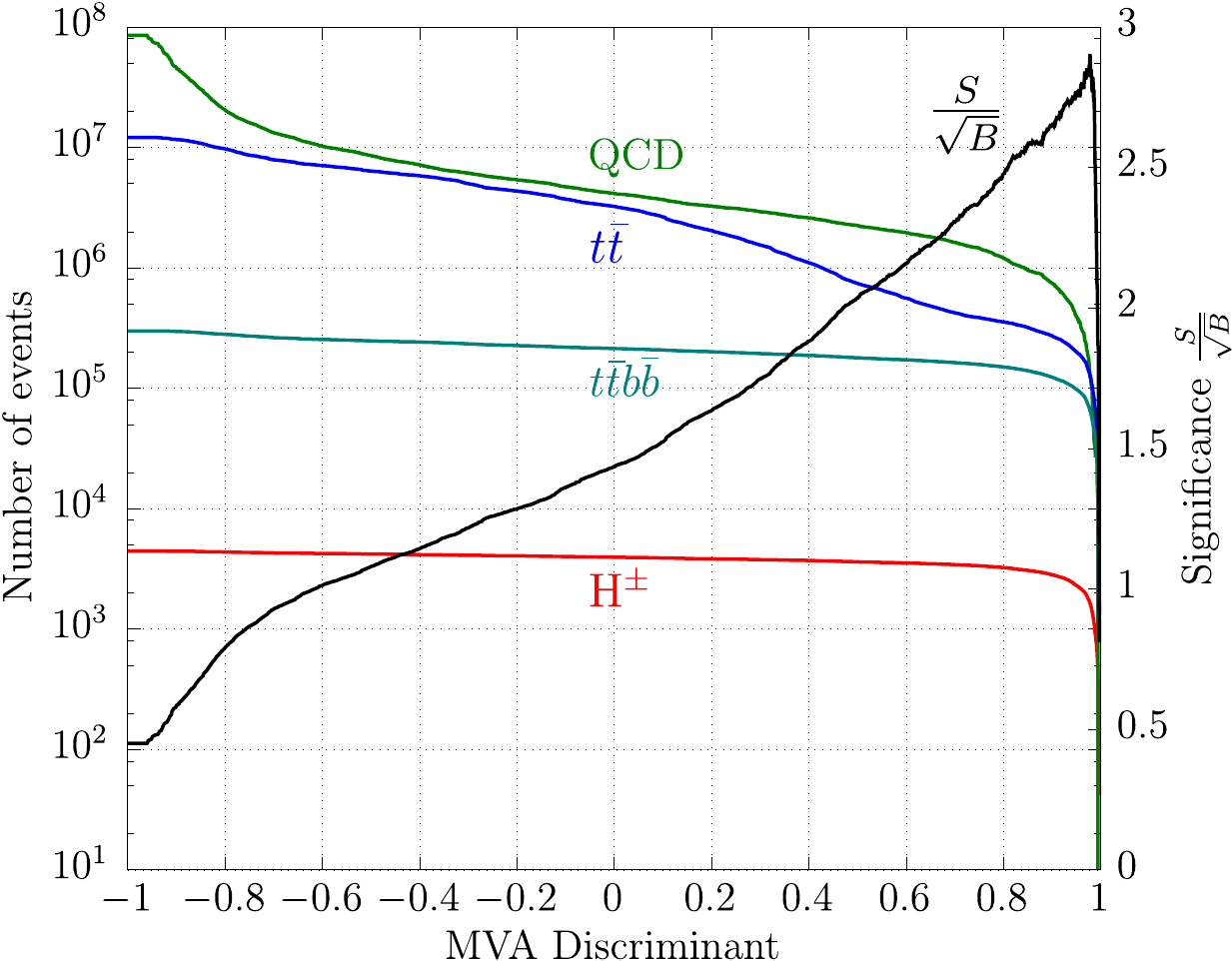}
    \end{center}
\end{figure}
In Fig.~\ref{fig:Ddist}, the distribution of 
MVA output discriminator (D) with the number of events are presented, 
for signal events with ${\mch}=500$~GeV and backgrounds from QCD,
$t\bar t$ and $t\bar t b \bar b$, along with the significance
$S/\sqrt{B}$ for an integrated luminosity 300 $\invfb$.
Significance close to $3\sigma$ can be achieved with a selection
of the discriminator, $\text{D}>0.9$. With this cut on D, and for
integrated luminosity of 300 fb$^{-1}$, the number of
events turn out to be 2830 for signal, and 1140000 for total backgrounds, 
where 70\% contribution come from QCD. 
The selection of $D>0.9$ leads to a significance $\sim 2.65 \sigma$ 
which goes up more for higher luminosity options. 

\begin{figure}[H]

    \begin{center}
        \caption{Same as for Fig.~\ref{fig:Ddist}, but for leptonic signal.}
        \label{fig:BDTlep}
        \includegraphics[width=0.8\textwidth]{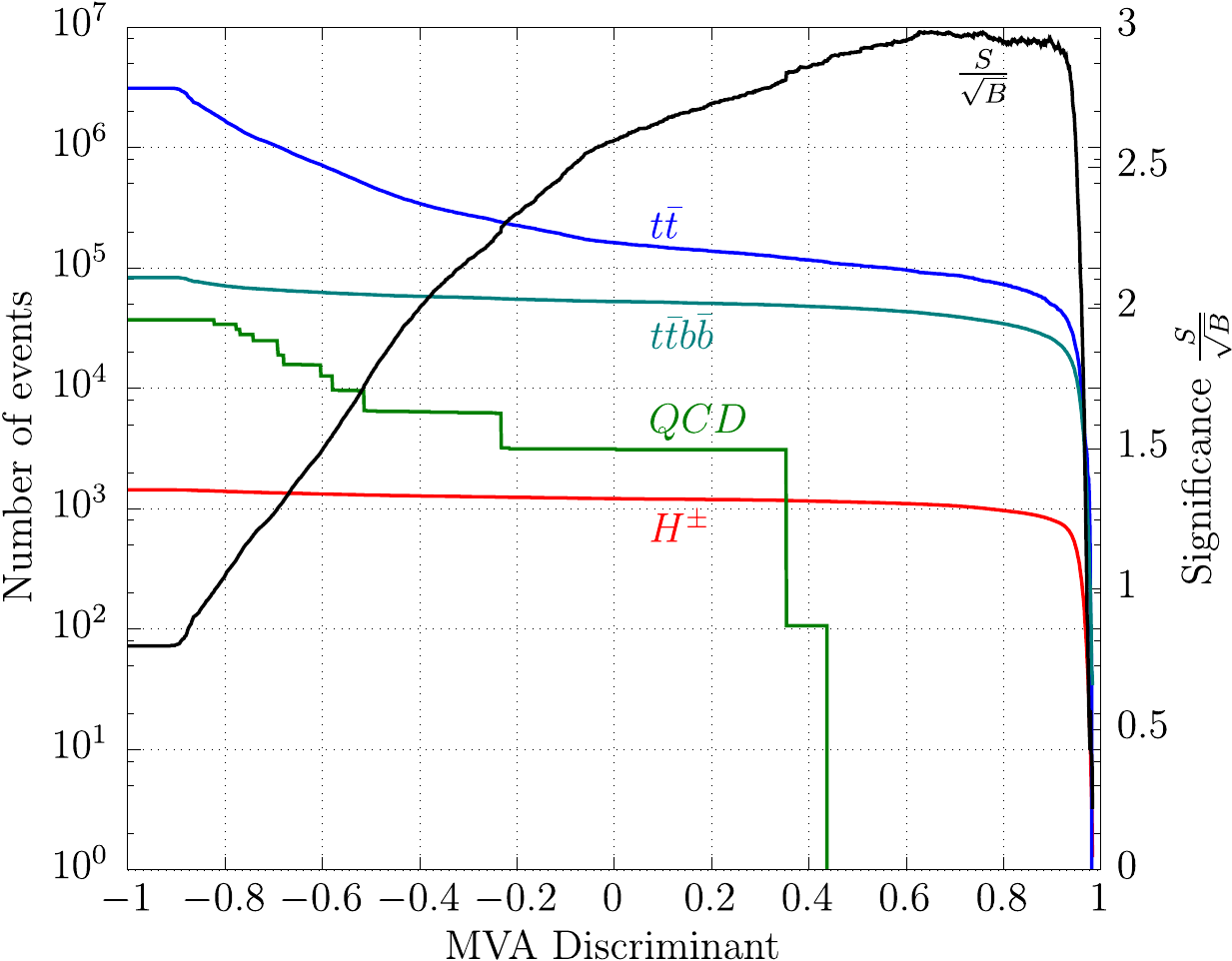}
    \end{center}
\end{figure}

Unlike the hadronic case, in the leptonic signal final 
state (see Fig.~\ref{fig:BDTlep}),
the dominant background appears to be due 
to $t \bar t$ production. A cut on BDT output $\text{D}>0.9$ leads to 
a significance of about 3$\sigma$ for ${\cal L}=300 \text{ fb}^{-1}$.
The study is extended upto the 1000~GeV mass of the charged Higgs.

Signal significances are presented for both hadronic and leptonic final 
state(in parenthesis) in Table.\ref{tab:SbyRootB} for three masses of 
charged Higgs and for three integrated luminosity options. 
Remarkably, using MVA technique a significant improvement in 
sensitivity for both hadronic and
leptonic signal is achieved.
This table suggests that 
in hadronic channel
the charged Higgs boson of mass
upto $\sim$ 500 GeV can be probed with a reasonable sensitivity,
much better than the obtained using simple cut based analysis as shown in, 
Table~\ref{tab:HadSbyB}.

\begin{table}[H]
    \begin{center}
        \caption{Significances for hadronic (leptonic) final states and for three luminosity options, in SUSY motivated Type II model with $\tan\beta=30$.}
        \label{tab:SbyRootB}
        \begin{tabular}{|c|cccc|}
            \hline 
            &  & $S/\sqrt{B}$  &  & \tabularnewline
            ${\mch}(GeV){\rightarrow}$  & 300  & 500  & 800  & 1000 \tabularnewline
            \hline 
            ${\cal L}=300$ fb$^{-1}$ & $6.1$ $\left(5.2\right)$ & $2.7$ $\left(2.94\right)$ & $0.61$ $\left(0.96\right)$ & $0.22$ $\left(0.39\right)$\tabularnewline
            ${\cal L}=1000$ fb$^{-1}$  & $11.0$ $\left(9.5\right)$ & $4.8$ $\left(5.4\right)$ & 1.1 $\left(1.7\right)$ & $0.40$ $\left(0.71\right)$\tabularnewline
            ${\cal L}=3000$ fb$^{-1}$  & $19.1$ $\left(16.5\right)$ & $8.4$ $\left(9.3\right)$ & 1.9 $\left(3.0\right)$ & $0.70$ $\left(1.2\right)$\tabularnewline
            \hline 
        \end{tabular}
    \end{center}
\end{table}
The Table~\ref{tab:SbyRootB} shows that the signature of charged Higgs of 
mass around 800 GeV is observable in leptonic channel for 3000~$\invfb$ luminosity option 
unlike the hadronic final state. For lower range of masses($\sim$ 500 GeV) 
signal is feasible even for 1000~$\invfb$ luminosity option. 

The results presented in Table~\ref{tab:SbyRootB} 
correspond to the SUSY motivated Type II model.
However, the signal cross sections for other classes of 2HDM can be obtained 
out of these estimated values simply 
by rescaling the couplings and appropriately multiplying
{\Brhtb}. The significances for all four types of models are presented 
for both both hadronic and leptonic(within the parenthesis) 
in Table~\ref{tab:SbyBCombAllTypeTb30} and ~\ref{tab:SbyBCombAllTypeTb3},  
corresponding to $\tan\beta=30$ and 3 respectively. 
Table~\ref{tab:SbyBCombAllTypeTb30} suggests that for 
high $\tan\beta$ scenario, discovery potential of charged Higgs in the context 
of Type II and Type IV model is quite promising for masses upto around 
600-700 GeV, however, due to little increase($\sim$ 20\%) of Br$\left( \Dchtb \right)$, 
sensitivity is better for Type IV model. Because of the suppressed coupling
of charged Higgs with top and bottom quarks, for high $\tan\beta$ scenario,
the signal sensitivity is very poor for both Type I and III model.    
However, for low 
$\tan\beta$ scenario(\ref{tab:SbyBCombAllTypeTb3}), results suggest that
discovery potential is quite promising for this kine of model parameter space.
Interestingly significances corresponding to all types of model,  
are found to be almost same for a given mass and 
luminosity option. It can be attributed to the fact, that in all classes of 
2HDM, the dominant part of charged Higgs couplings proportional to 
$m_t\cot\beta$ which results the same significances.    

\begin{table}
    \begin{center}
        \caption{
                Sensitivity($S/{\sqrt{\rm B}}$) for hadronic (leptonic)
                signal events corresponding to all 4 types of 2HDM, and $\tan\beta=30$,
                $\sin\left(\beta-\alpha\right)=1$.
        }
        \label{tab:SbyBCombAllTypeTb30} %
        \begin{tabular}{|c|c|c|c|c|c|}
            \hline
            $\mch$ (GeV)  & ${\cal L}$ (in fb$^{-1}$)  & Type I  & Type II  & Type III  & Type IV \tabularnewline
            \hline
            & 300  & 0.043 $\left({0.037}\right)$  & {$7.4$} $\left({6.4}\right)$  & {$0.0007$} $\left({0.0006}\right)$  & $9.270$ $\left({7.999}\right)$\tabularnewline
            300  & 1000  & {$0.08$} $\left({0.07}\right)$  & {$13.5$} $\left({11.6}\right)$  & {$0.001$} $\left({0.001}\right)$  & {$16.925$} $\left({14.603}\right)$\tabularnewline
            & 3000  & {$0.14$} $\left({0.12}\right)$  & {$23.3$} $\left({20.1}\right)$  & {$0.002$} $\left({0.002}\right)$  & {$29.316$} $\left({25.294}\right)$\tabularnewline
           \hline
            & 300  & {$0.017$} $\left({0.019}\right)$  & {$3.1$} $\left({3.4}\right)$  & {$0.0004$} $\left({0.0004}\right)$  & {$3.621$} $\left({4.017}\right)$\tabularnewline
            500  & 1000  & {$0.031$} $\left({0.034}\right)$  & {$5.6$} $\left({6.2}\right)$  & {$0.0007$} $\left({0.0009}\right)$  & {$6.611$} $\left({7.335}\right)$\tabularnewline
            & 3000  & {$0.053$} $\left({0.059}\right)$  & {$9.8$} $\left({10.8}\right)$  & {$0.001$} $\left({0.001}\right)$  & {$11.451$} $\left({12.704}\right)$\tabularnewline
            \hline
            & 300  & {$0.004$} $\left({0.006}\right)$  & {$0.71$} $\left({1.1}\right)$  & {$0.0001$} $\left({0.0002}\right)$  & {$0.823$} $\left({1.294}\right)$\tabularnewline
            800  & 1000  & {$0.007$} $\left({0.011}\right)$  & {$1.3$} $\left({2.0}\right)$  & {$0.0002$} $\left({0.0003}\right)$  & {$1.502$} $\left({2.363}\right)$\tabularnewline
            & 3000  & {$0.01$} $\left({0.02}\right)$  & {$2.2$} $\left({3.5}\right)$  & {$0.0003$} $\left({0.0005}\right)$  & {$2.601$} $\left({4.093}\right)$ \tabularnewline
           \hline
        \end{tabular}
    \end{center}
\end{table}

\begin{table}
    \begin{center}
        \caption{Same as Table.\ref{tab:SbyBCombAllTypeTb30}, but with $\tan\beta=3$.\\}
        \label{tab:SbyBCombAllTypeTb3}
        \begin{tabular}{||c|c|c|c|c|c||}
            \hline 
            $\mch$ (GeV)  & ${\cal L}$ (in fb$^{-1}$)  & Type I  & Type II  & Type III  & Type IV \tabularnewline
            \hline 
            & 300  & {$4.3$} $\left({3.7}\right)$ & {$4.3$} $\left({3.7}\right)$ & {$4.3$} $\left({3.7}\right)$ & {$4.3$} $\left({3.7}\right)$\tabularnewline
            300  & 1000  & {$7.8$} $\left({6.7}\right)$ & {$7.9$} $\left({6.8}\right)$ & {$7.9$} $\left({6.8}\right)$ & {$7.8$} $\left({6.7}\right)$\tabularnewline
            & 3000  & {$13.5$} $\left({11.7}\right)$ & {$13.7$} $\left({11.8}\right)$ & {$13.7$} $\left({11.8}\right)$ & {$13.5$} $\left({11.7}\right)$\tabularnewline
            \hline 
            & 300  & {$1.7$} $\left({1.9}\right)$ & {$1.7$} $\left({1.9}\right)$ & {$1.7$} $\left({1.9}\right)$ & {$1.7$} $\left({1.9}\right)$\tabularnewline
            500  & 1000  & {$3.1$} $\left({3.4}\right)$ & {$3.1$} $\left({3.5}\right)$ & {$3.1$} $\left({3.5}\right)$ & {$3.1$} $\left({3.4}\right)$\tabularnewline
            & 3000  & {$5.3$} $\left({5.9}\right)$ & {$5.4$} $\left({6.0}\right)$ & {$5.4$} $\left({6.0}\right)$ & {$5.3$} $\left({5.9}\right)$\tabularnewline
            \hline 
            & 300  & {$0.39$} $\left({0.62}\right)$ & {$0.40$} $\left({0.63}\right)$ & {$0.40$} $\left({0.63}\right)$ & {$0.39$} $\left({0.62}\right)$\tabularnewline
            800  & 1000  & {$0.72$} $\left({1.1}\right)$ & {$0.73$} $\left({1.1}\right)$ & {$0.73$} $\left({1.1}\right)$ & {$0.72$} $\left({1.1}\right)$\tabularnewline
            & 3000  & {$1.2$} $\left({2.0}\right)$ & {$1.3$} $\left({2.0}\right)$ & {$1.3$} $\left({2.0}\right)$ & {$1.2$} $\left({2.0}\right)$\tabularnewline
            \hline
        \end{tabular}
    \end{center}
\end{table}
The discovery potential of charged Higgs of mass 300 GeV, 
in the $\bar {t}{ b}$ decay channel is quite promising even 
for 300 $\invfb$ luminosity option at 13 TeV energy.
However, for higher masses, e.g. for $\mch=500$ GeV, one needs high
luminosity options such as 1000 $\invfb$ and more. This study shows that 
for higher masses $\sim$ 1000~GeV, it is very hard to achieve 
better signal sensitivity even for high luminosity option.

\begin{figure}[H]
	\begin{center}
		\caption{Discovery region for a given luminosity(in $\invfb$) 
			options in the context of SUSY(Type II) model for hadronic(left) and
			leptonic(right) final state at $\sqrt{s}=13$ TeV.}
		\label{fig:region}
		\includegraphics[width=0.49\textwidth]{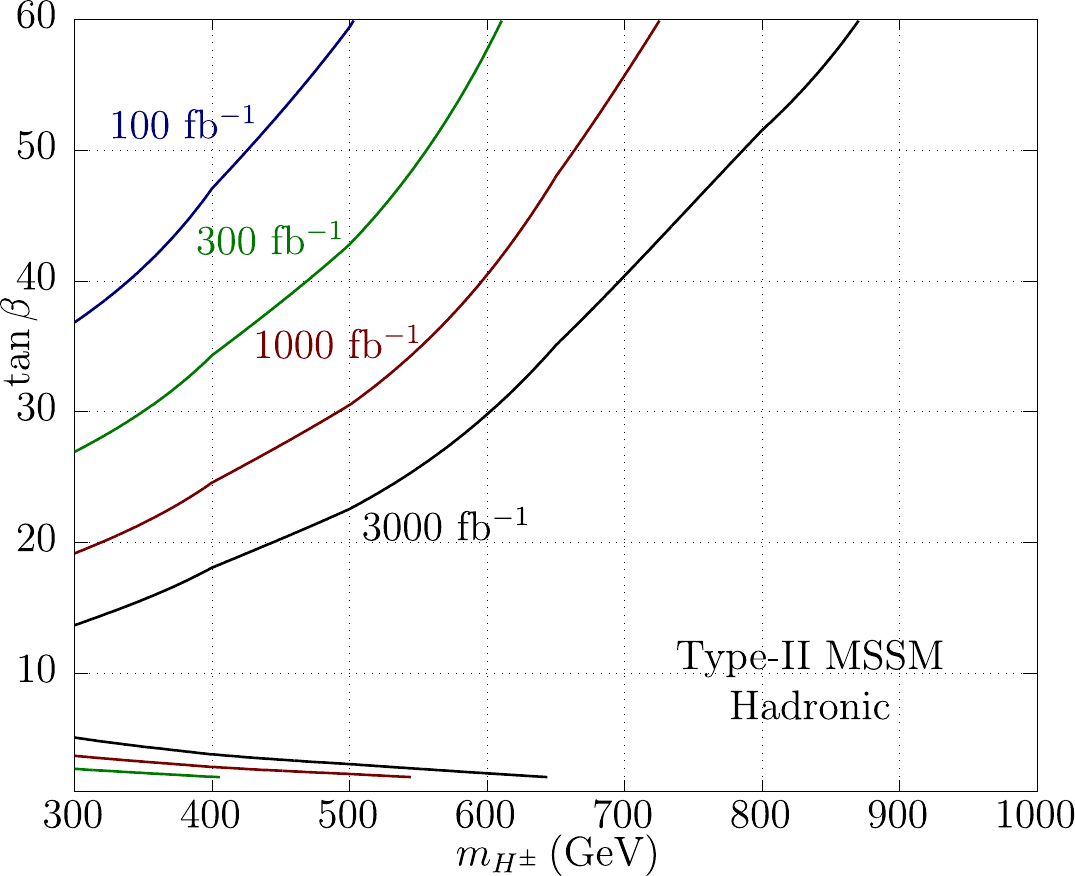}
		\includegraphics[width=0.49\textwidth]{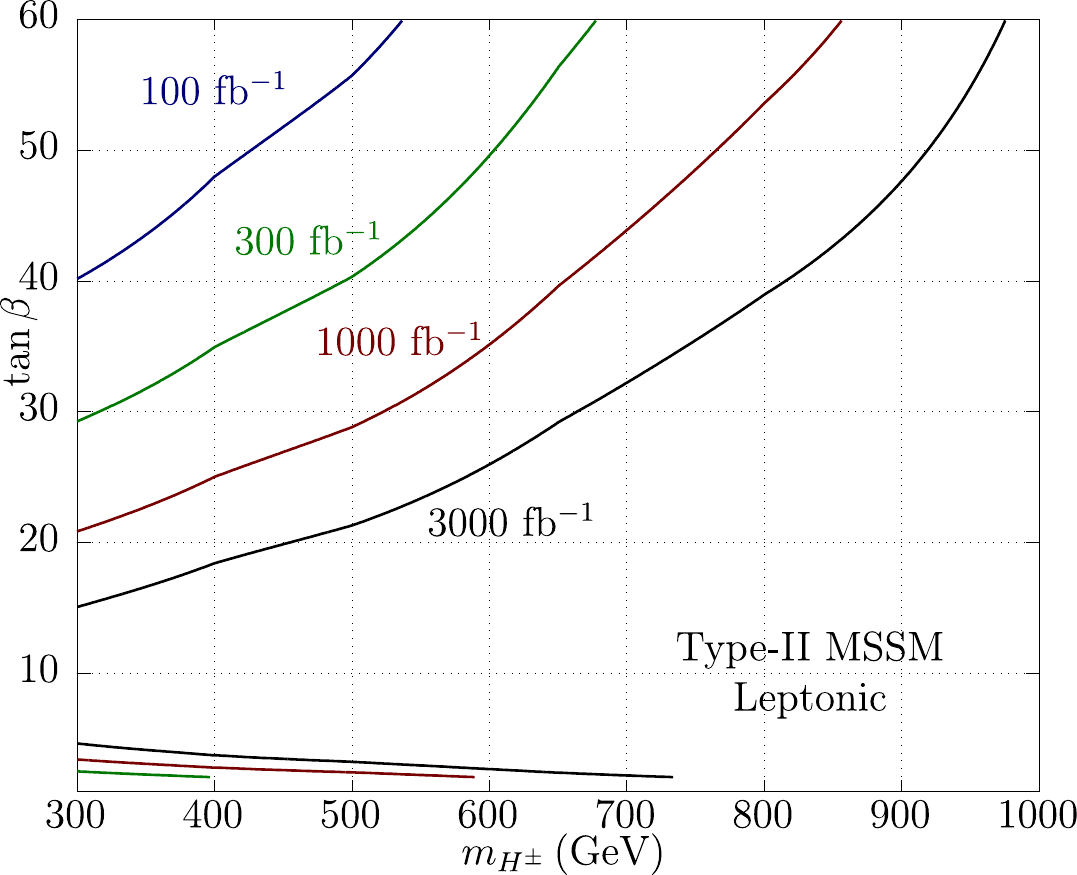} 
	\end{center}
\end{figure}
 
Finally, in Fig.~\ref{fig:region} the 
discovery region is presented  
in the $\mch - \tan\beta$ plane 
in the context of SUSY motivated Type II model, requiring a 5$\sigma$
significance for a given luminosity options, as shown in the figure.
Contours show that the minimum value of $\tan\beta$ required to discover 
charged Higgs of given mass mass at $5\sigma$ level for a given 
luminosity option. 
The parameter space above the contours are discoverable
corresponding to that luminosity option at $\sqrt{s}=$13~TeV energy.
In hadronic channel, even for high luminosity option, it is very hard to 
find charged Higgs of mass beyond 850~GeV. On the other hand, for the 
leptonic final state, charged Higgs can be explored almost 
upto $\mch \sim$1~TeV with high luminosity. Discovery regions below the 
contours for much lower
$\tan \beta$ are also shown with three luminosity options. For a given
$\mch$, the lowest $\tan\beta$ corresponds to 3000$\invfb$ and then 
decreases to 1000$\invfb$ and 300$\invfb$ for other two lines respectively.

It is to be noted that while calculating the signal significance, the 
uncertainties of the background are not taken into account. The estimation
of systematics in the background evaluation is currently out of the
scope for the present analysis. 
However, due to the tiny signal size in comparison to background events, i.e
with less purity, the impact of systematic uncertainties is expected to be
severe. It can be understood by evaluating the significance as, 
S/${\sqrt{B + (\delta B)^2}}$, where $\delta$ stands for the level
of uncertainties. For instance, corresponding to moderate range of
charged Higgs masses, and for about 20\% uncertainties in background
estimation, the signficances go down drastically, for both hadronic
and leptonic case.
For heavier mass range, $\mch\sim$ 800 GeV, the impact of systematics to 
significance is not that severe
due to less number of background events.
Clearly, in order to achieve a reasonable significance to discover the 
charged Higgs for the intermediate mass range, one needs to
perform the background estimation as precisely as possible.

\section{Summary}
\label{summary}
In this study, we explore the detection prospect of the charged 
Higgs boson for the heavier mass range at the LHC in Run 2 experiments 
with the center of mass energy,
$\sqrt{s}=13$~TeV, within the framework of generic 2HDM.
A very brief discussion of 2HDM is presented in
order to set up model framework to carry out the analysis. It is observed,
that in all classes of 2HDM, the {\Brhtb} always 
the dominant one, except in Type III model where it holds only for
lower range of $\tan\beta \left ( < 10 \right )$. It is to be noted that
the other decay modes, such as $\chh \to {W}^{\pm} \phi$ $\left (\phi \in \left\{ h, H, A \right\} \right)$ also open up
with a large Br once the condition $\sin(\beta-\alpha)=1$ is relaxed, as 
discussed in Sec.\ref{modelref}.
The charged Higgs boson production cross section at
$\sqrt{s}=13$ TeV are computed in both 4FS and 5FS mechanisms, and finally
the matched values are presented for a few representative choices of $\mch$.
In the context of SUSY motivated Type II model, the matched cross
sections vary from ${\cal O}$(100~fb) to
${\cal O}$(10~fb) for the range of $\mch \sim 300 - 1000$ GeV corresponding to
large values of $\tan\beta$, and found to be less for other classes 
of 2HDM. The signature of charged Higgs is analyzed for the final state 
consisting of a reconstructed charged Higgs mass and extra $b$-jets plus
an additional reconstructed top quark for hadronic events, while 
in leptonic events, a lepton is required without reconstruction of
second top.
The jet substructure technique is used to tag moderately boosted top
quark from heavier charged Higgs decay in order to avoid
re-combinatorial problem while reconstructing the charged Higgs mass.
The MVA method is employed including inputs from HepTopTagger to tag 
topjets. A better top tagging efficiency with lower mis-tagging rate 
is achieved in comparison to the result obtained using only 
default HepTopTagger. 
The detailed simulation is performed for signal, and the main 
dominant irreducible SM backgrounds from the top quark
pair production and QCD. The cut based analysis predicts very poor 
signal sensitivity even for high luminosity options.  
However, for
lower mass range of charged Higgs,
${\mch}\sim 300$ GeV, one can expect a modest sensitivity for 
3000 $\invfb$ luminosity option. 
In order to improve the signal significance, the analysis is carried out using
the techniques of BDT within the framework of TMVA.
Several kinematic variables are constructed to train BDT.
Remarkably, MVA analysis yields substantial improvement in 
signal significance.
For example, this MVA based analysis shows that 
with ${\cal L}=1000$ $\invfb$, the signature of  charged Higgs
boson for the mass range $\sim 300-700$  GeV can be probed for both hadronic
and leptonic channel. For more higher luminosity option, such as
3000~$\invfb$, the discovery reach of {\mch} can be extended up to 
$\sim$ 800~GeV for hadronic final state, where as for leptonic case,
it can be extended further, up to almost 1 TeV for high values of 
$\tan\beta$.
In Fig.~\ref{fig:region}, the discovery potential of charged Higgs boson
are presented in the $\mch- \tan\beta$ plane for a few integrated 
luminosity options. This figure  indicates that the discovery reach 
corresponding to leptonic final state is better than the hadronic signal
case.   
Simply scaling the charged Higgs couplings with fermions, and then the 
production cross sections, 
we present signal significances for all classes of 2HDM for three
representative choices of $\mch$ and two 
values of $\tan\beta=30$ and 3.
The results show that for high $\tan\beta=30$ scenario,
it is difficult to achieve any detectable signal sensitivity, 
except for Type II and Type IV models.
However, for low $\tan\beta \left( =3 \right)$ case,
the signal of charged Higgs for the mass 
range $\sim 300-600$ GeV seems to be detectable
with a $\sim 3 \sigma$
sensitivity for leptonic final state with ${\cal L}=1000\invfb$.
Indeed, it is hard to discover the signal of the charged Higgs 
boson of mass beyond 800~GeV for low $\tan\beta$ scenario, even for higher
luminosity options. 
Definitely, to probe charged Higgs boson of very mass, more than 800 GeV,
one needs very high energy option, such as 100 TeV hadron  
Collider\cite{Mangano:2017tke}.  

\section{Acknowledgments:}
The authors are thankful to Abhishek Iyer and Rickmoy Samanta for joining 
this project at the earlier stage. We used feynrules~\cite{Alloul:2013bka} 
model file uploaded by authors of \cite{Degrande:2015vpa}, and also contacted
C. Degrande (one of the author) regarding running b-quark masses 
in the model. MG acknowledges support from CERN theory division where the last phase of the work 
is done. 

\bibliography{paperv2.bib}
\bibliographystyle{utphys.bst}
\end{document}